\documentclass[a4paper,11pt]{article}
\pdfoutput=1 

\usepackage{jinstpub} 
\usepackage{graphicx}

\title{\boldmath Numerical estimation of discharge probability in GEM-based detectors}

\author[a,b,1]{Prasant Kumar Rout,\note{Corresponding author.}}
\author[a,b]{R. Kanishka,}
\author[a,b]{Jaydeep Datta,}
\author[a,b]{Promita Roy,}
\author[c]{Purba Bhattacharya,}
\author[a,b]{Supratik Mukhopadhyay,}
\author[a,b]{Nayana Majumdar,}
\author[a,b]{Sandip Sarkar}

\affiliation[a]{Saha Institute Of Nuclear Physics, \\1/AF Saltlake, Kolkata 700064, India}
\affiliation[b]{Homi Bhabha National Institute, Training School Complex, \\Anushaktinagar, Mumbai 400094, India}
\affiliation[c]{Department of Physics, University of Calcutta, \\92 A.P.C Road, Kolkata 700009, West Bengal, India}

\emailAdd{prasantrout7@gmail.com}

\abstract{Discharge probability in GEM-based gaseous detectors has been numerically estimated using an axisymmetric hydrodynamic model.
Initial primary charge configurations in the drift region, obtained using Heed and Geant4, are found to have significant effect on the subsequent evolution of detector response.
Simulation of energy resolution has been performed to establish the capability of the hydrodynamic model to capture statistical nature of the experimental situation.
Finally, single and triple GEM configurations exposed to alpha sources have been simulated to estimate discharge probability which have been compared with available experimental data.
Despite the simplifying and drastic assumptions in the numerical model, the comparisons are encouraging.}        


\arxivnumber{2103.12849} 

\begin{document}
\maketitle
\flushbottom

\section{Introduction}
\label{sec:Intro}
Micro-Pattern Gaseous Detectors, in particular the Gas Electron Multiplier(GEM) ~\cite{Sauli1997}, are widely used in the high luminosity experiments such as COMPASS~\cite{Abbon}, LHCb~\cite{LHCb2001}, ALICE~\cite{ALICE1999}, CMS~\cite{CMS2015} and TOTEM~\cite{Bozzo, Bagliesi}. 
Over the years a lot of effort has been put together in the research and development of this technology for improving the performance of GEM-based detectors for long term operation in the mentioned high rate experiments. 
These detectors have been used both as tracking and triggering detectors due to their excellent spatial resolution of $\sim$30 $\mu m$ and time resolution in the nanosecond range \cite{Titov:2007fm}. 
Apart from the position and time resolution performances of these detectors, parameters like radiation hardness, aging resistance, high rate capability and high voltage stability against electrical discharges are of crucial importance, especially for long term operation of different experiments. 
Electrical dischrages are among the most important threats to all the experiments that may lead to instability and irreversible damages to the detectors.
As a result, many studies have been performed to investigate formation of discharge in GEM-based detectors \cite{Abbas, Bencivenni:2003qs, Bencivenni:2002ij, Croci}.

There are several factors that determine the transition from avalanche to discharge formation.
Some of the more prominent ones are the number of primaries, electric field configuration and the transport, amplification, attachment properties of the gas mixture.
There is general agreement that the transition from avalanche to discharge is dependent on a critical charge density, termed as the Raether limit ~\cite{Raether1964}.
This limit is found to be dependent on the gas mixture, size of the typical detector structure and is reported to be within the range (5-9)$\times$10$^{6}$ electrons~\cite{Gasik:2017uia} for argon(Ar)-based and neon(Ne)-based mixtures.
Large charge densities produced in the drift region by highly-ionizing particles like alpha makes the detector prone to discharges and hence significantly alter the stability of the detector.
It has also been observed in ~\cite{Gasik:2017uia} that if the charge deposit occurs close to the GEM hole, discharge probability is higher.
The use of highly-ionizing alpha source on single, double and triple GEM detectors with Ar-based gas mixtures was also reported in ~\cite{Bachmann:2000az}.
The relation of discharge probability with number of primaries in the drift region indicates that these detectors are likely to face increasing difficulty as the luminosity of the experiments increase.

In multistage GEM-based detectors, the gas amplification process is shared among the GEM foils and this can be used to delay the onset of discharges in such configurations.
According to ~\cite{Bressan:1998uu}, the occurrence of discharge can be prevented by decreasing the gas amplification from the GEM foil facing the incident radiation to the foil facing towards the anode electrode so that it does not reach the critical avalanche size. 
This is possible by lowering the applied potential difference across different GEMs successively.
This lowering scheme of potential difference across GEM foils has been experimentally illustrated in \cite{Bachmann:2000az} and such an asymmetric distribution of potential differences was successfully used in COMPASS~\cite{Altunbas:2002ds,Ketzer:2002ii}.
Recently, the triple GEM detectors being used to upgrade the Muon system of CMS~\cite{CMS2015} also implements similar distribution of amplification fields in different GEM layers.
This distribution, however, has adverse effect on the ion back-flow suppression which is of importance for a GEM-based TPC~\cite{Bhattacharya:2017yaj,daLuz:2018ngj,Ball:2014qaa}.

Computer simulation of discharges is known to be a difficult problem because of the strongly nonlinear nature of the phenomenon.
While several attempts have been made in different contexts ~\cite{Laurence,M,Jannis,Sentman}, as far as we know, only few attempts have been made to model discharges in gaseous ionization detectors ~\cite{Gasik:2017uia,Datta:2020whg}, despite its acknowledged importance in various experimental situations.
As a result, estimation of discharge probability using purely numerical approach has remained beyond the reach of the detector community. 
In~\cite{Gasik:2017uia}, in addition to experimental studies, a numerical model based on Geant4 \cite{Agostinelli:2002hh} has been used to estimate the discharge probability for single GEMs with success.
However, aspects related to the transport and amplification were ignored in the model and a combination of computational and experimental approach was utilized instead.
Moreover, a discharge was defined by comparing the number of accumulated charges inside a single GEM hole with a predefined critical charge limit that was estimated separately.
The dynamics of processes leading to the charge accumulation was not followed in detail and the transition from avalanche to discharge was not simulated.
It is easy to appreciate that a relatively more independent and reliable numerical estimation of discharge probability will be of significant help to different experimental groups, especially if it can be used to analyze multi-GEM detectors as well.
In this work, attempts have been made to address this issue by developing a fast hybrid numerical model to simulate the entire chain of events starting from primary ionization to formation of avalanche, or discharges in single and triple GEM detectors.
Attempts have been made to estimate discharge probability in such detectors being operated under different experimental conditions using the same numerical model.

The proposed model is based on a recently-developed hydrodynamic model ~\cite{Rout:2020bjz} which could simulate the transition from avalanche to discharges in multiple GEM-based detectors reasonably well. 
It was found, in agreement with several earlier numerical models ~\cite{Gasik:2017uia, Fonte_RPC, avalanches_discharge,Fillipo}, that the presence of high charge densities in the close vicinity of GEM foil gives rise to the space charge dominated discharges driven by ions in GEM detectors.
In the present paper, we have extended the 2D axisymmetric hydrodynamic model to simulate the response of single and triple GEM detectors exposed to different radiation sources, as well as to estimate discharge probabilities due to heavily-ionizing alpha particles.
The simulated results are obtained with the Ar$-$carbon dioxide(CO$_{2}$) gas mixture in volume proportions 70$-$30.
In order to emulate the statistical nature of different processes determining the evolution of charge transport within the detector, fluctuations in primary electron configurations have been considered in fair detail and provided as an input to the hydrodynamic model.
The rest of the present paper is organized as follows.
Section \ref{sec:Sim} describes numerical model adopted for estimating the primary ionization following which the transport and amplification of charges in the gas volume is discussed.
The details of the hydrodynamic model were described in detail in ~\cite{Rout:2020bjz} and are briefly mentioned here.
Section \ref{sec:Results} contains energy resolution estimates that is used to establish the capability of the proposed model to reproduce statistical processes.
The method adopted for estimating discharge probability and the simulated values of the discharge probability are also presented in this section and compared with experimental results from \cite{Bachmann:2000az}.
Finally, section \ref{sec:Conclusion} contains the summary of the work and few concluding remarks.                   
   
\section{Simulation details}
\label{sec:Sim} 
The numerical studies have been carried out in two steps:
(1) Generation of primaries,
(2) Transport and amplification of primaries.
The first step is carried out using Monte Carlo models, while the latter uses a hydrodynamic model, resulting into a hybrid numerical model for the entire process.
Here, both the numerical steps are described in brief.

\subsection{Generation of primaries}
\label{subsec:EventGen}
This initial distribution of primary electrons plays an important role in determining the subsequent evolution of the charged particles.
Estimation of such initial configurations is conveniently implemented using Monte-Carlo techniques.
Thus, in this part of the computation, number and location of primaries generated within the detector gas volume are estimated using Geant4 and Heed ~\cite{Heed}, depending on the type of source used.
For example, the former was used to estimate the primaries generated due to an alpha source, as described in some detail below.

In the experimental arrangement described in ~\cite{Bachmann:2000az}, a collimated beam of alpha particles with an angular opening of $\pm$ 30$^{\circ}$ from an $^{241}$Am alpha source placed on a thin (3.5 $\mu$m) mylar window external to the semi transparent drift cathode was used for measurement of discharge probability. 
Following this description, a three-dimensional simulation geometry has been created in Geant4, that closely mimics the experimental arrangement. 
A collimated beam of alpha particles with an angular opening of $\pm$30$^{\circ}$ passing through a thin mylar window of thickness 3.5 $\mu$m has been injected into the gas volume of the numerical model.
This active gas volume, made up of Ar$-$CO$_{2}$ in proportions 70$-$30 by volume, was considered to be a three-dimensional box of dimensions 5 cm along x, 5 cm along y and 20 cm along z direction.
Geant4 follows particle transport in prefixed, or automatically decided steps with possibility of an interaction after each step.
The step-lengths are computed according to the cross-sections of physics processes to be considered for a particular simulation.
The possible physics processes are maintained in different physics lists among which the user selects a few  depending on their suitability for simulating a given physics problem.
For the present studies, physics lists like EmPenelope, EmLivermorePhysics, Photo Absorption Ionization (PAI) and PAI-photon were used to simulate the particle interations with the active detector material.
The position and energy deposit of each individual Geant4 hit were extracted from the simulation for each of the alpha tracks in the gas volume.
Results corresponding to such event generation exercises will be presented in section \ref{subsec:Primary}.
 
\subsection{Transport and amplification of primaries}
\label{subsec:HydroModel}
In ~\cite{Rout:2020bjz}, the hydrodynamic approach for transport and amplifrication of primaries has been presented in great detail.
The mathematical model of this approach is based on the mass transport of diluted chemical species in a background solvent gas mixture.
The electrons and ions are considered as charged fluids. 
One of the major advantages of this hydrodynamic approach is the natural inclusion of space charge effects throughout the evolution of the charged species.
The entire mumerical model has been built in the framework of a commercially available Finite Element Method (FEM) Package COMSOL MultiPhysics \cite{COMSOL}.
The Transport of Dilute Species (TDS) module of COMSOL was specifically used to model the transport of charged fluids/species.
The drift-diffusion equations shown in \ref{eq:BTE1} $-$ \ref{eq:BTE5} govern the charge transport in the mathematical model.
\begin{equation}
\label{eq:BTE1}
\centering
{ \frac{\partial{c_{e}}}{\partial{t}} + {\vec{\nabla}}\cdot({-D_{e}\vec{\nabla}{c_{e}} + \vec{u_{e}}}{c_{e}}) = R_{e}}
\end{equation}
\begin{equation}
\label{eq:BTE2}
\centering
{ \frac{\partial{c_{i}}}{\partial{t}} + {\vec{\nabla}}\cdot({-D_{i}\vec{\nabla}{c_{i}} + \vec{u_{i}}}{c_{i}}) = R_{i}}
\end{equation}
\begin{equation}
\label{eq:BTE3}
\centering
{{R_{e}} =  {R_{i}} = {S_{e}} + {S_{ph}}}
\end{equation}
\begin{equation}
\label{eq:BTE4}
\centering
 {{S_{e}} = ({\alpha{(\vec{E})}} - {\eta{(\vec{E})}}){\mid{\vec{u_{e}}}}\mid{c_{e}}}
\end{equation}
\begin{equation}
\label{eq:BTE5}
\centering
S_{ph} = \xi QE_{gas}\mu\Psi_{0}
\end{equation}
where $c_{e}$, $R_{e}$, $\vec{u_{e}}$ and $D_{e}$ indicates the concentration, rate of production, drift velocity and diffusion coefficients for electron fluid respectively. 
On the otherhand, $c_{i}$, $R_{i}$, $\vec{u_{i}}$ and $D_{i}$ represents the concentration, rate of production, drift velocity and diffusion coefficients for ion fluid respectively.
The electrons and ions are produced in the same rate from the gas amplification processes as given in the equation \ref{eq:BTE3}. 
Townsend ionization given by $S_{e}$ and Photo-ionization mechanism represented by $S_{ph}$ serve as the source of production for electrons and ions in the gas volume. 
$\alpha$ and $\eta$ represent the first Townsend and Attachment coefficients of electron for the Townsend ionization term ($S_{e}$) in equation \ref{eq:BTE4} respectively.
Various electron transport coefficients namely, $\alpha$, $\eta$, $\vec{u_{e}}$ and $D_{e}$ are evaluated as a function of electric field in the gas mixture Ar-CO$_{2}$ using \cite{Magboltz}.

Photons released as a result of the de-excitation of a gas mixture portion with a higher ionisation energy, 
These photons can ionise the other gases in the mixture with lower ionisation energies.
For example, in Ar-CO$_{2}$ gas mixture, Ar is used as the couting gas, while CO$_2$ serves as a photon quencher.  
Photons are emitted as an excited Ar atom gets de-excited.
Since Ar has a higher ionisation energy than CO$_2$, the de-excitation of the excited Ar atoms ionises the CO$_2$ molecules via photo absorption.  
This is referred to as Photo-ionization, and it occurs in parallel with Townsend ionisation. 
Photo-ionization serves as the photon feedback in the model.
The photon feedback provides the contribtion of UV photons generated in the gas volume towards electron-ion production.
The photo absorption coefficient and UV photon flux produced in the detector volume are denoted by the terms $\mu$ and $\Psi_0$ in the equation \ref{eq:BTE5} respectively. 
The quantum efficiency of a gas mixture due to photon absorption is given by $QE_{gas}$ and $\xi$ denotes the proportion of excited states that can ionise the gas.  
Photon transport in the gas volume can be described using the following equation, as demonstrated by Capeill{\`{e}}re et al. in \cite{Capeillere2008}.
\begin{equation}
\label{eq:RTE_Photon}
\centering
{\vec{\nabla}(-c{\vec{\nabla}}{\Psi_{0}}) + a{\Psi_{0}} = f}
\end{equation}
where $c$ = $\frac{1}{3\mu_{all}}$, \hskip 2pt $f$ = ${\delta}$ $S_{e}$, \hskip 2pt $a$ = $\mu_{all}$.   
${\delta}$ represents the number of excited neutral Ar atoms per ionizion.
Using the photo absorption cross section from \cite{Sahin:2014haa}, the average photo-absorption coefficient ($\mu_{all}$) of CO$_2$ was determined, taking into account all the excited states of Ar. 
Equation \ref{eq:RTE_Photon} corresponds to the diffusion-like approximation to the photon flux(${\Psi_{0}}$) field.
The "Coefficient Form Partial Differential Equation" module of COMSOL is used to model photon propagation in the current simulation model. 

We have calculated the photo-absorption coefficient ($\mu$) using only those excited states of Ar that has energy greater than the ionisation energy of CO$_2$ because all photons absorbed by CO$_2$ would not be able to ionize further. 
From the data \cite{NiIST_Ar}, we also calculated the fraction of these excited states of Ar and used it as a multiplicative factor $\xi$ in $S_{ph}$ as shown in equation \ref{eq:BTE5}. 
In table \ref{tab:table1}, we have compiled a list of all the photo-ionization parameters that have been used. 
\begin{table}[h!]
  \begin{center}
    \begin{tabular}{|c|c|}
      \hline
      \textbf{Variable} & \textbf{Value} \\
      \hline
       $\delta$ & 10$^{-4}$ \\
       \hline
       QE$_{\rm gas}$ & 10$^{-4}$  \\
       \hline
       $\mu_{\rm all}$ & 981/cm  \\
      \hline
      $\mu$ & 831/cm \\
      \hline
      $\xi$ & 0.133 \\
      \hline
    \end{tabular}
    \caption{A set of variables used for photo-ionization and photon transport}
    \label{tab:table1}
  \end{center}
\end{table} 

The charge transport strongly depends on the electric field configuration and gets modified due to the effect of space charges.
The "Electrostatics" module of COMSOL calculates the electric field at each time step optimised by COMSOL. 
The following equations \ref{eq:Field}-\ref{eq:poisson} are utilized to compute the electric potential(V) and field $(\vec{E})$.  
\begin{equation}
\label{eq:Field}
\centering
  {{\vec{E}} = -\vec{\nabla}{V}}
\end{equation}
\begin{equation}
\label{eq:poisson}
\centering
{{\vec{\nabla}}.\vec{D} = {\rho_{v}}} 
\end{equation}
where, $\vec{D}$ represents the electric displacement vector, 
$\vec{D} = \epsilon_{0} \epsilon_{r} \vec{E}$, $\epsilon_{0}$ being the permittivity of vacuum or air and $\epsilon_{r}$ denotes the relative permittivity of the material used in the detector geometry.
The volume space charge density in the gas volume has been incorporated in the model and is calculated using the equation \ref{eq:space_charge_density}.
\begin{equation}
\label{eq:space_charge_density}
\centering 
{\rho_{v} = \frac{Q_{e}}{\epsilon_{0}}(c_{i} - c_{e})}
\end{equation}
Where, $\rho_{v}$ corresponds to the volume space density and $Q_{e}$ denotes the charge of an electron.

Optimization of three Dimesional (3D), 2D axisymmetric and 2D model geometries of GEM, on the basis of electric field configurations and computational speed and resources have been explored to find the optimum model for describing the charge dynamics in these detectors. 
Out of these three model configurations, 3D model was found to be realistic and provides correct field strength and true picture of charge dynamics in GEM based detectors.
However, it is computationally expensive.
2D model consists of straight channels and was considered to be inconsistent, because of incorrect estimation of field strength. 
The 2D axisymmteric model on the otherhand has a central hole with the symmetry axis passing through its center.
Many off-centre holes are also present in the peripheral region which are represented as circular channels instead of holes as shown in figure 1(e) of \cite{Rout:2020bjz}.
The central hole of the model provides a correct field distribution as that of the 3D model, whereas the off-centre holes have a field discrepancy of nearly 20\%. 
The central hole contributes substantially towards the Townsend avalanche and discharge formation, whereas the off-centre holes contribute less because of reduced field. 
In order to take into account the contribution of off-centre holes, an approximate scale factors \cite{Rout:2020bjz} has been derived by utilizing the charge sharing information and field values in the off-centre holes which improves the effective gain estimate in the model.
Thus 2D axisymmetric model has been found to be suitable in the present work as it is reasonably fast and provides correct field map in the central hole for study of the charge dynamics of avalanche and discharges.  

In the present model, a discharge occurs in the 2D axisymmetric gas volume when the total number of electrons generated exceeds a limit value.
This value depends on the gas mixture and the typical detector structure as mentioned in section \ref{sec:Intro}.
The threshold value has been obtained to be 8$\times$10$^{5}$ electrons for single GEM and 2$\times$10$^{6}$ electrons for double and triple GEM detectors \cite{Rout:2020bjz} respectively.
These values are supported by \cite{Gasik:2017uia}, where similar values of discharge limit (5$\times$10$^{6}$ electrons) have also been reported for Ar-CO$_{2}$ (70-30) gas mixtures in GEM-based detectors. 

\begin{figure}[htbp]
\centering
\includegraphics[width=\linewidth]{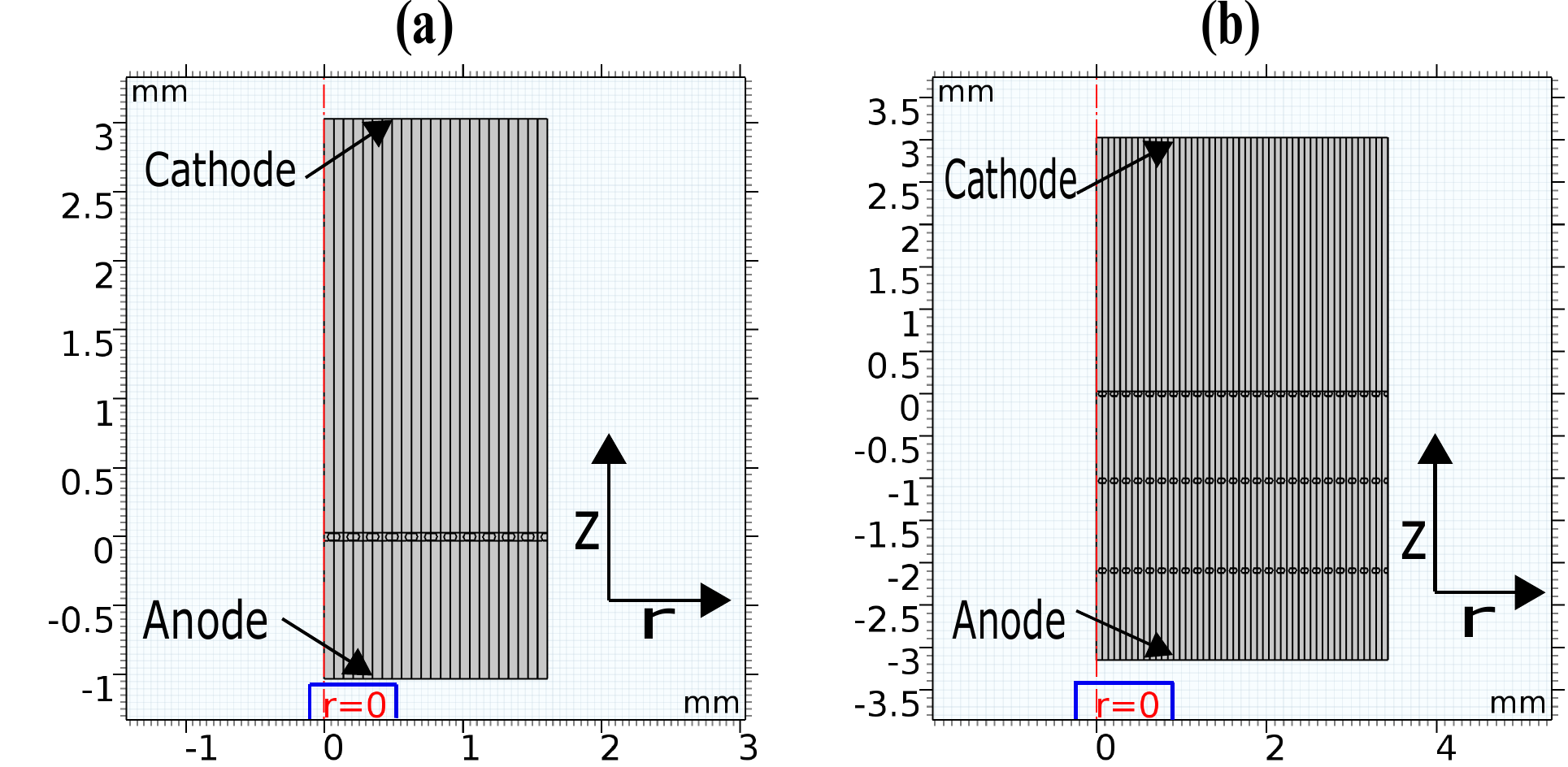}
\caption{2D axisymmetric geometry of (a) single GEM and (b) triple GEM configurations}
\label{fig:ModelGeometry} 
\end{figure}

The 2D axisymmetric model of single and triple GEM configurations utilized in the present work are shown in figures \ref{fig:ModelGeometry} (a) and (b), respectively.
The electric field configuration and the parameters of the detector geometries follow the description of ~\cite{Bachmann:2000az} and are shown in table \ref{tab:table2}. 
\begin{table}[h!]
  \begin{center}
    \begin{tabular}{|c|c|c|c|c|c|}
    \hline
      \textbf{GEM Structures} & \textbf{Range} & \textbf{Drift Field} & \textbf{Transfer  Field} & \textbf{Induction Field} & \textbf{Gap} \\
      & $\Delta{V}_{\rm GEM}$(V) & (kV/cm) & (kV/cm) & (kV/cm) & (mm)\\
      \hline
      Single GEM & 500 - 515 & 2 & -   & 3.5 & 3:1\\
      \hline
      Triple GEM & 385 - 400 & 2 & 3.5 & 3.5 & 3:1:1:1\\
      \hline
    \end{tabular}
    \caption{Electric potential, field configuration and geometrical parameters for single and triple GEM structures used in the simulation}
    \label{tab:table2}
  \end{center}
\end{table}

One important point to note is the fact that hydrodynamic models are deterministic and influence of statistical fluctuations are usually not implemented when adopting such an approach.
However, discharge probability and many other important properties of gaseous ionization detectors are, by nature, statistical and it is important to incorporate this aspect into the final estimates of the numerical model.
This has been a major challenge throughout this investigation and, in order to satisfy this requirement, fluctuations arising in primary generation were used as inputs to different executions of the hydrodynamic model and the final estimate of the statistical quantities were derived from individual results of each of these simulations.
It should be mentioned here that the execution time for each of the single GEM configuration for avalanche (discharge) mode operations are around 15 (90) minutes, while for a triple GEM configuration, it was close to 45 (210) minutes respectively, on a Xeon-based workstation having eight cores and 64GB RAM.
According to several test executions, the time taken is similar on generation 7 i5 desktop machine having 8GB RAM.
In order to confirm that the proposed hybrid numerical model works with dependable accuracy, energy resolutions of single and triple GEM detectors have been estimated and will be presented next in section ~\ref{subsec:Gain}.

\section{Results and Discussions}
\label{sec:Results}

\subsection{Energy resolution estimate}
\label{subsec:Gain}
The energy resolution of GEM-based detectors depends on several important factors that include the statistical nature of primary ionization, fluctuations in the avalanches, changes in temperature and pressure, asymmetry in the detector geometry, intrinsic noise in the readout electronics and many more \cite{Jonathan}. 
As a result, the mean energy (E) of the photon is found to be smeared to a gaussian of standard deviation ($\sigma_{E}$) instead of a sharp delta function peak \cite{Jonathan}.
The energy resolution can be obtained as the Full Width Half Maximum (FWHM) over E as shown in equation \ref{eq:E1}.
\begin{equation}
\label{eq:E1}
\centering
{ \frac{\sigma_{E}}{E} = \sqrt{\frac{F}{\overline{n}} + (\frac{\sigma_{\overline{G}}}{\overline{G}})^{2} + (\Delta_{0}^2)}}
\end{equation}

Here, F is the gas-dependent Fano factor and its value for Ar-CO$_{2}$(70-30) gas mixture has been found out to be $\sim$0.221 \cite{Purba}.
$\overline{G}$ is the mean gas gain, $\sigma_{\overline{G}}$ being the standard deviation of $\overline{G}$.
The quantity $\overline{n}$ represents the average number of primaries found to be 211 using \cite{Heed} and $\Delta_{0}$ indicates 
the various systematic variations in the detector, such as effects of fluctuations like foil asymmetry, temperature and pressure, etc.
In the present version of the model, $\Delta_{0}$ has been considered to be zero. 

The energy resolution has been estimated by incorporating gain fluctuations in the model. 
The following three sources of fluctuation have been incorporated:
\begin{itemize}
  \item Fluctuations in the number of primaries.
  \item Variation of spatial distribution of primary seed cluster.
  \item Variation of mean z position of primary seed cluster in the drift region.
\end{itemize}  
Realistic configurations of the seed clusters have been set up by analysing the number and location information of the primaries from the $^{55}$Fe radiation source obtained using \cite{Heed}.
The fluctuation of the number of primaries have been considered from 190-230.
Four mean z position of primary seed cluster have been considered in the drift region, such as 0.250 mm, 0.5 mm, 0.75 mm and 0.9 mm. 
The spatial distributions of the primary seed cluster along X, Y and Z directions have been presented in figure 4 of~\cite{Rout:2020bjz}.
The mean cluster spread of the primary seed cluster was obtained to be 0.132 mm in radial (r) and 0.154 mm along z direction.
However, many possibilities of cluster spread can result from these distributions.
In the present work, 1$\sigma$ of the mean cluster spread has been chosen as its variation by fitting a gaussian function to the cluster spread distribution of primaries shown in figure 4 of~\cite{Rout:2020bjz}. 
The fit results a $\sigma$ of 0.071 mm in r and 0.08 mm along z direction.
Thus, in addition to the mean cluster spread, other four extreme cases of cluster spread configuration have been obtained by addition and subtraction of 1$\sigma$ from the mean cluster spread along r and z directions and are shown in table \ref{tab:table3}.
Each of the five cases of cluster spread are considered as initial seeds, can be located at the above four mean z positions in the drift region.
\begin{table}[h!]
  \begin{center}
    \begin{tabular}{|c|c|c|c|}
    \hline
      \textbf{Number of Cases} & \textbf{Description} & \textbf{Cluster spread in r} & \textbf{Cluster spread in z} \\
      & & (mm) & (mm)\\
      \hline
      Case 1 & mean & 0.132 & 0.154 \\
      \hline
      Case 2 & enlarged & 0.203 & 0.234 \\
      \hline
      Case 3 & shrunk & 0.061 & 0.074 \\
      \hline
      Case 4 & r-elongated & 0.203 & 0.074 \\
      \hline
      Case 5 & z-elongated & 0.061 & 0.234 \\
      \hline
    \end{tabular}
    \caption{Five cases of cluster spread for $^{55}$Fe radiation source}
    \label{tab:table3}
  \end{center}
\end{table}

The position and shape of seed cluster in the drift gap affect the drift and diffusion of electrons and ions that lead to different charge sharing patterns in the GEM holes. 
The contribution of the primary electrons to the central hole of an axisymmetric geometry reduces significantly with increasing distance of the primary seed cluster along the z direction towards the cathode in the drift region.
Starting with a number of primaries and a certain mean z position, the evolution of charged fluid have been simulated in the five different cases of cluster spread variation.
Similar reduction to the central hole contribution is observerd for cases with large cluster spread.
The axisymmetric estimates of effective gain values have been obtained for all these possible configurations and have been scaled to that of the appropriate 3D values by using scale factors derived as described in ~\cite{Rout:2020bjz}.  

\begin{figure}[htbp]
\centering
\includegraphics[width=0.4\linewidth]{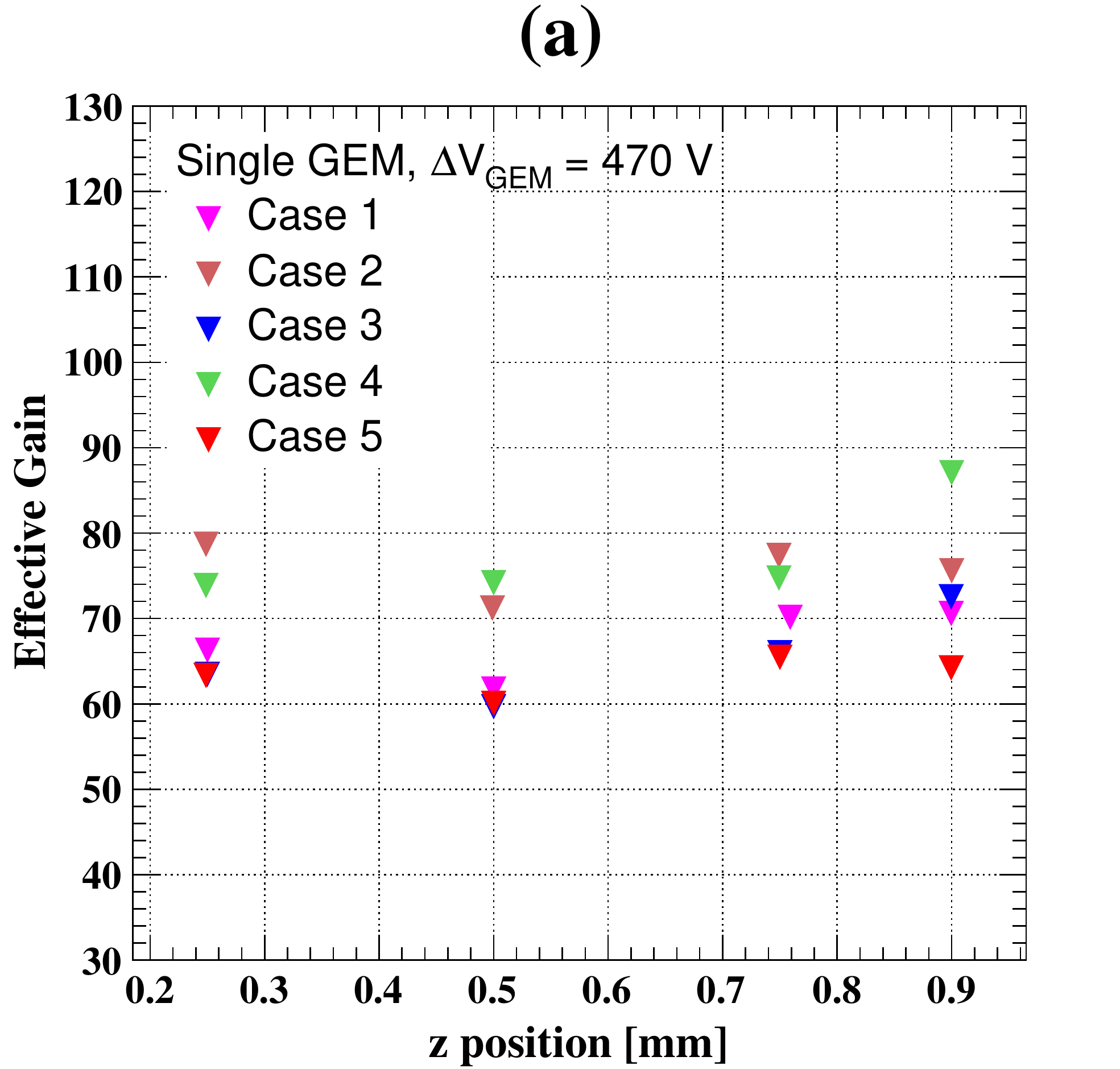}
\includegraphics[width=0.4\linewidth]{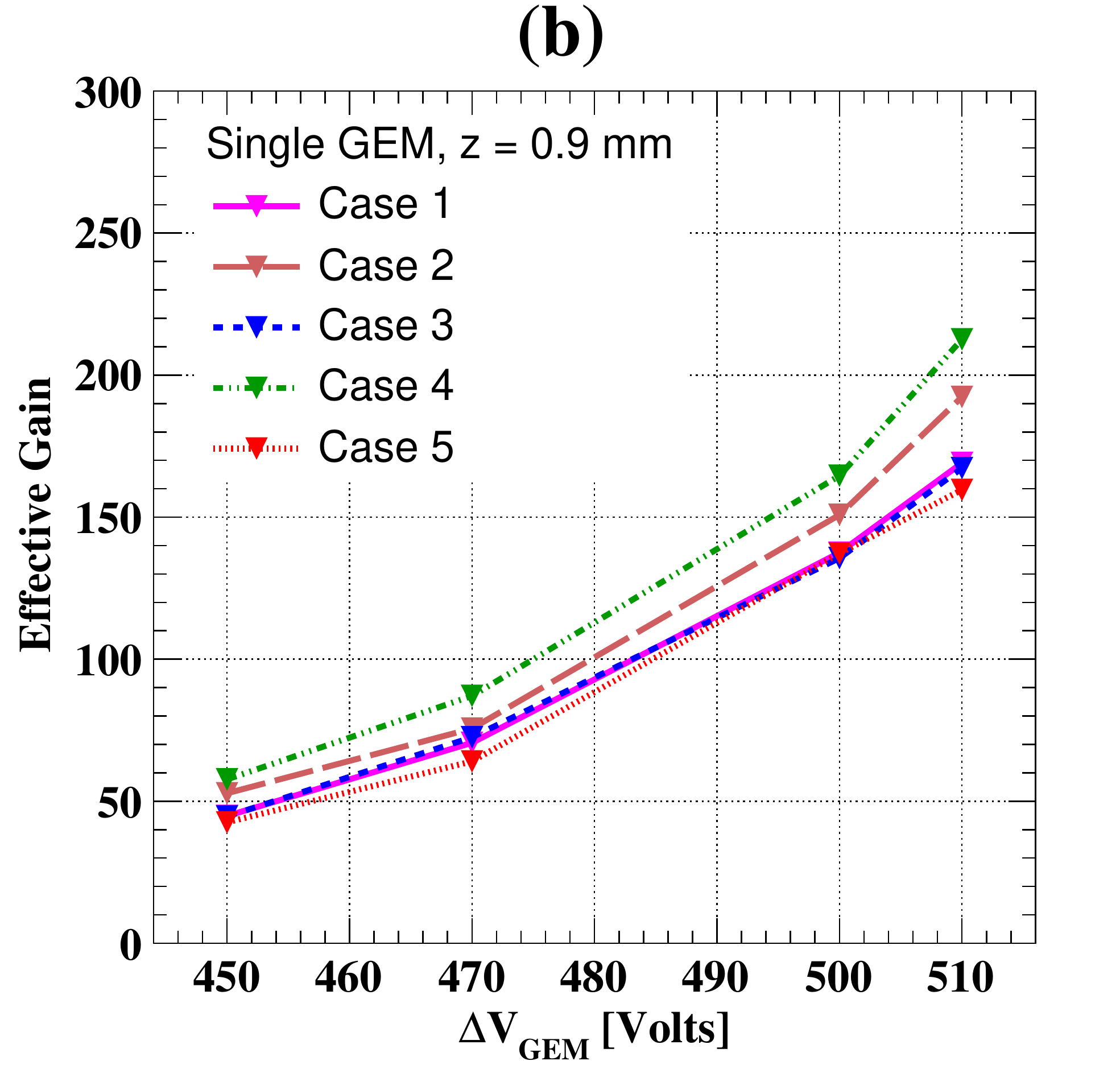}
\includegraphics[width=0.4\linewidth]{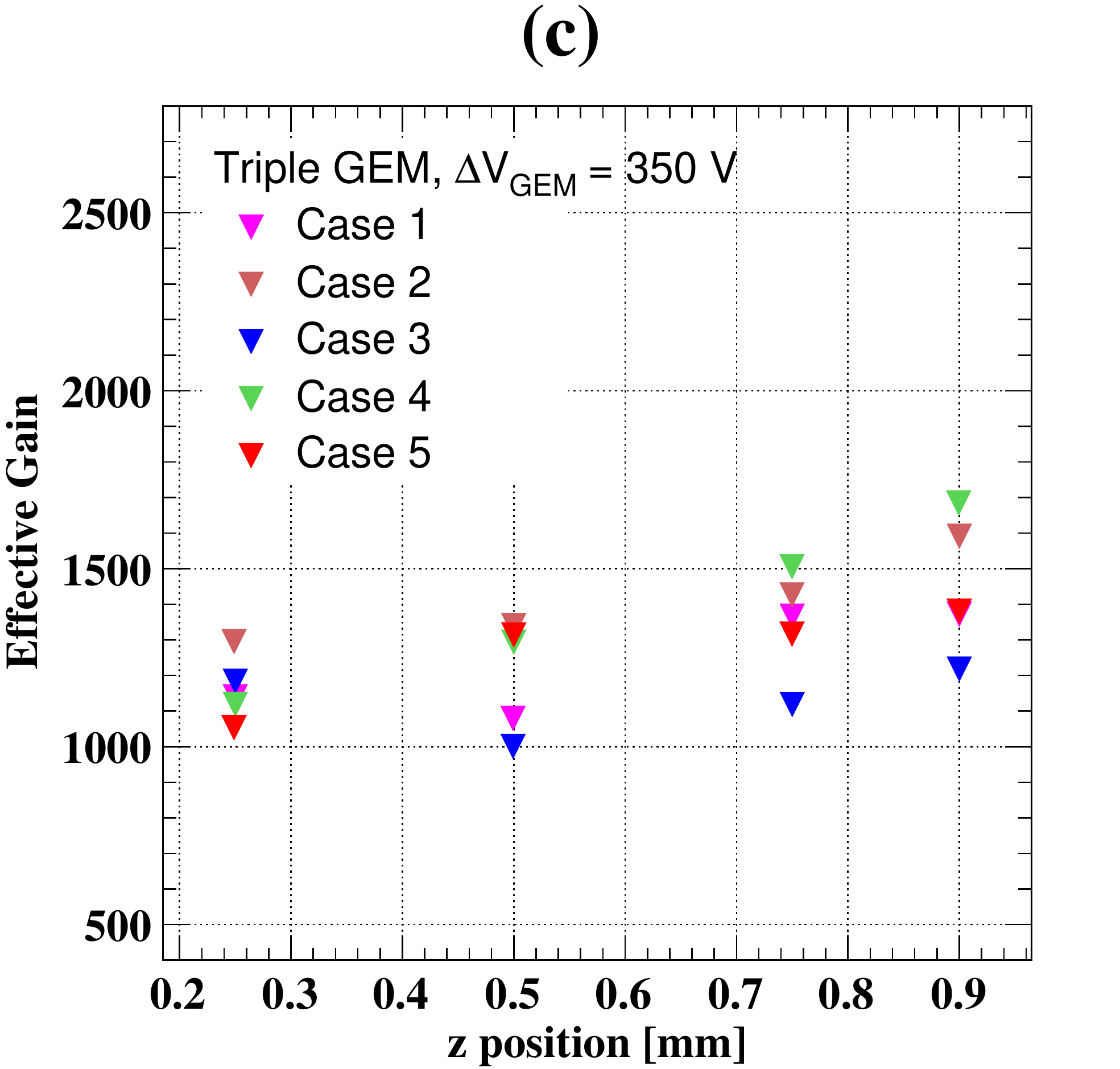}
\includegraphics[width=0.4\linewidth]{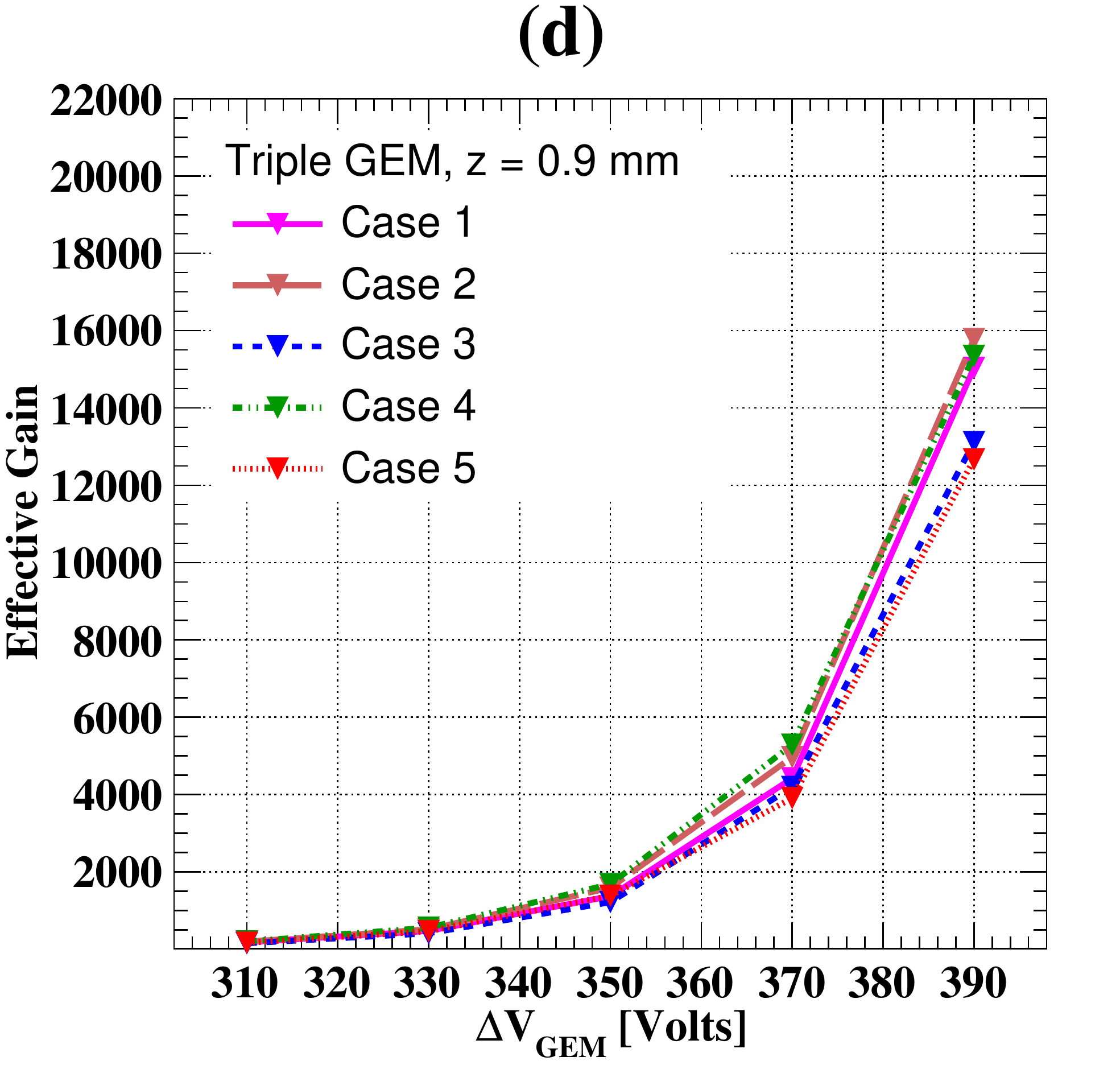}
\caption{(a) Numerical estimates of effective gain variation with different mean z positions of the primary seed cluster in a single GEM for five cases of cluster spread at $\Delta{V}_{GEM}$ = 470 V. (b) Effective gain with different applied voltages to the GEM foils for five different cases of seed cluster spread placed at z = 0.9 mm in the drift gap of single GEM. (c) Effective gain variation with different mean z positions of primary seed cluster for five different cases at $\Delta{V}_{GEM}$ = 350 V. (d) Effective gain with different applied voltages to the GEM foils in five different configurations of seed cluster placed at z = 0.9 mm in the drift gap for a triple GEM.}
\label{fig:GainVariation}
\end{figure}

Figure \ref{fig:GainVariation}(a) shows the numerical estimates of effective gain for five different cases of cluster spread with the mean z position of the  primary seed cluster for a single GEM with an applied voltage of $\Delta{V}_{GEM}$ = 470 V to the GEM foil.
The variation of effective gain with voltages applied to the GEM foil of a single GEM detector for different shapes of the primary seed cluster placed at a certain position (z = 0.9 mm) in the drift gap has also been shown in figure \ref{fig:GainVariation}(b).
Similarly, for these five different cases of seed cluster, the effective gain values of a triple GEM detector for a fixed $\Delta{V}_{GEM}$ = 350 V applied to the three GEM foils with different mean z positions of the seed cluster in the drift gap has been shown in figure \ref{fig:GainVariation}(c), while the variation of gain as a function of applied voltages to the three GEM foils of a triple GEM is shown in figure \ref{fig:GainVariation}(d).
It may be noted that both collection efficiency and scale factor have been incorporated while estimating the effective gain values.
It may be concluded that number, position and shape of the primary cluster have significant effect on the estimated effective gain.
The enlarged (case 2) and r-elongated (case 4) seeds tend to result into larger values of gain.
The rest of the configurations, namely, mean (case 1), shrunk (case 3) and z-elongated (case 5), give rise to similar values of gains.
Moreover, seeds located at larger distances from the first GEM towards the cathode in the drift region lead to slightly larger values of gain.

Further studies relating to the hydrodynamic predictions reveal that there are three reasons that leads to this difference in gain values as the location and size of the cluster is modified:
\begin{itemize}
\item{effect of changes in collection efficiency,}
\item{effect of space charge (a factor related to charge sharing),}
\item{effect of unphysical electric field differences between central hole and off-centre channels}.
\end{itemize}
Among these, the third and clearly wrong component has been neutralized to some extent through the use of the scale factor.
There is scope of significant improvement in the process of estimation of this factor but, for a fast simulator, the present simple approach seems to be reasonable.
The collection efficiency has been obtained using Garfield\cite{Garfield}-neBEM\cite{neBEM}-Magboltz framework.
We have initiated the study on the effect of space charge and its dependence on the initial conditions of the simulation.
The preliminary observations are the following:
\begin{itemize}
\item{Keeping all other parameters constant, enlarging the cluster size reduces the space charge effect: leads to an increase in effective gain,}
\item{Keeping all other parameters constant, the space charge effect increases as voltage is increased: leads to a suppression of the effective gain at larger gain values.}
\end{itemize}
Similar results are obtained for triple GEM structures in which electric field at each level is found to be influenced by the space charge accumulation.
However, since the applied voltage at each stage is small in comparison to the single GEM detector, the effect is prominent only on the GEM foil adjacent to the induction gap.

According to our present understanding, the first point is the reason why Case 2 and Case 4 values are greater than the other three cases.
In these two cases, the spread of the cluster in the radial direction is larger than in the other (Case 2 is the largest among all the cases).
As a result, the central hole has the least number of electrons and ions at any given moment.
Thus, the space charge effect is also less in these cases, leading to larger values of gain.
This is a very interesting observation and we hope to investigate the space charge dynamics in relation to charge sharing further in the near future.

\begin{figure}[htbp]
\centering
\includegraphics[width=0.4\linewidth]{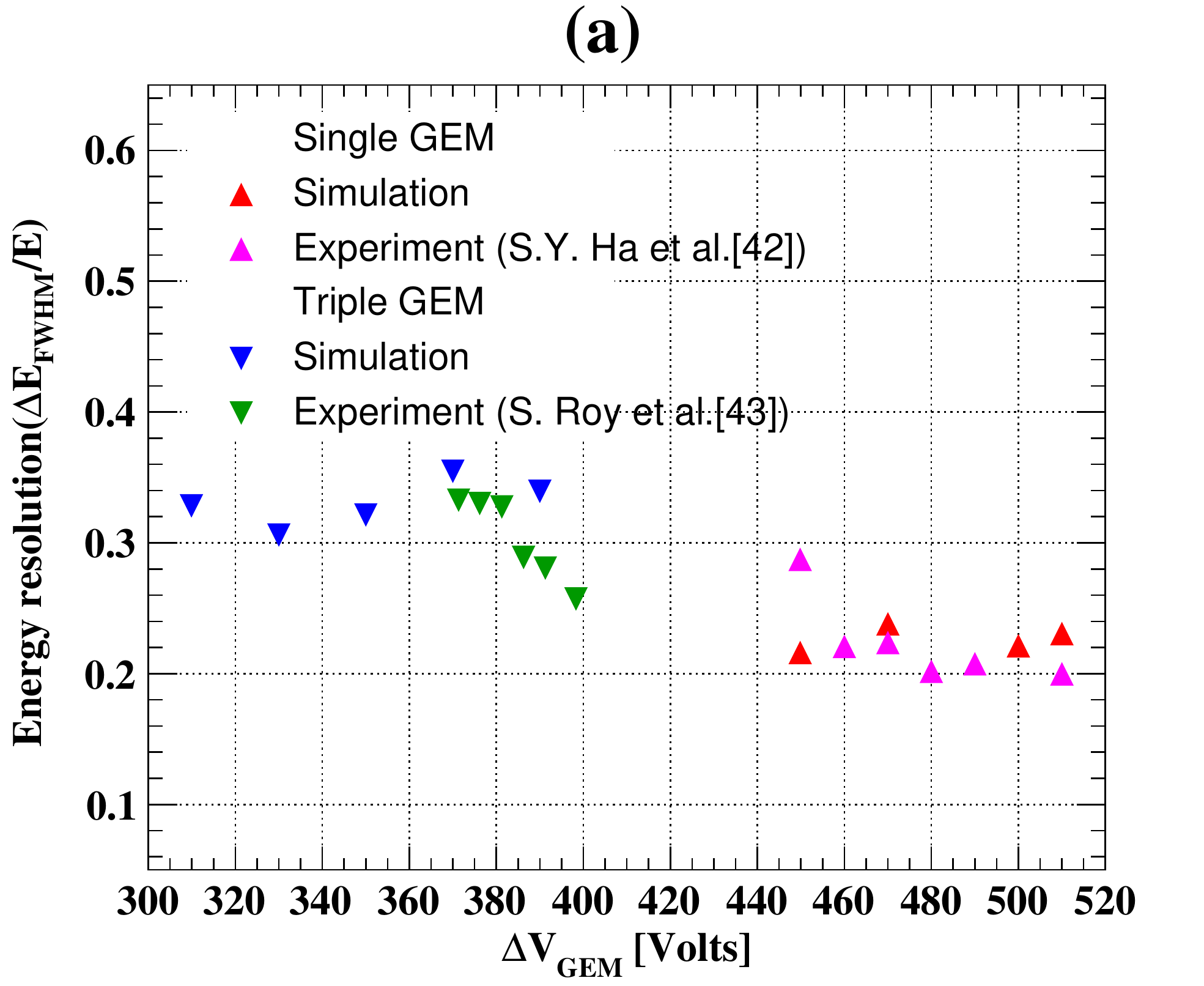}
\includegraphics[width=0.4\linewidth]{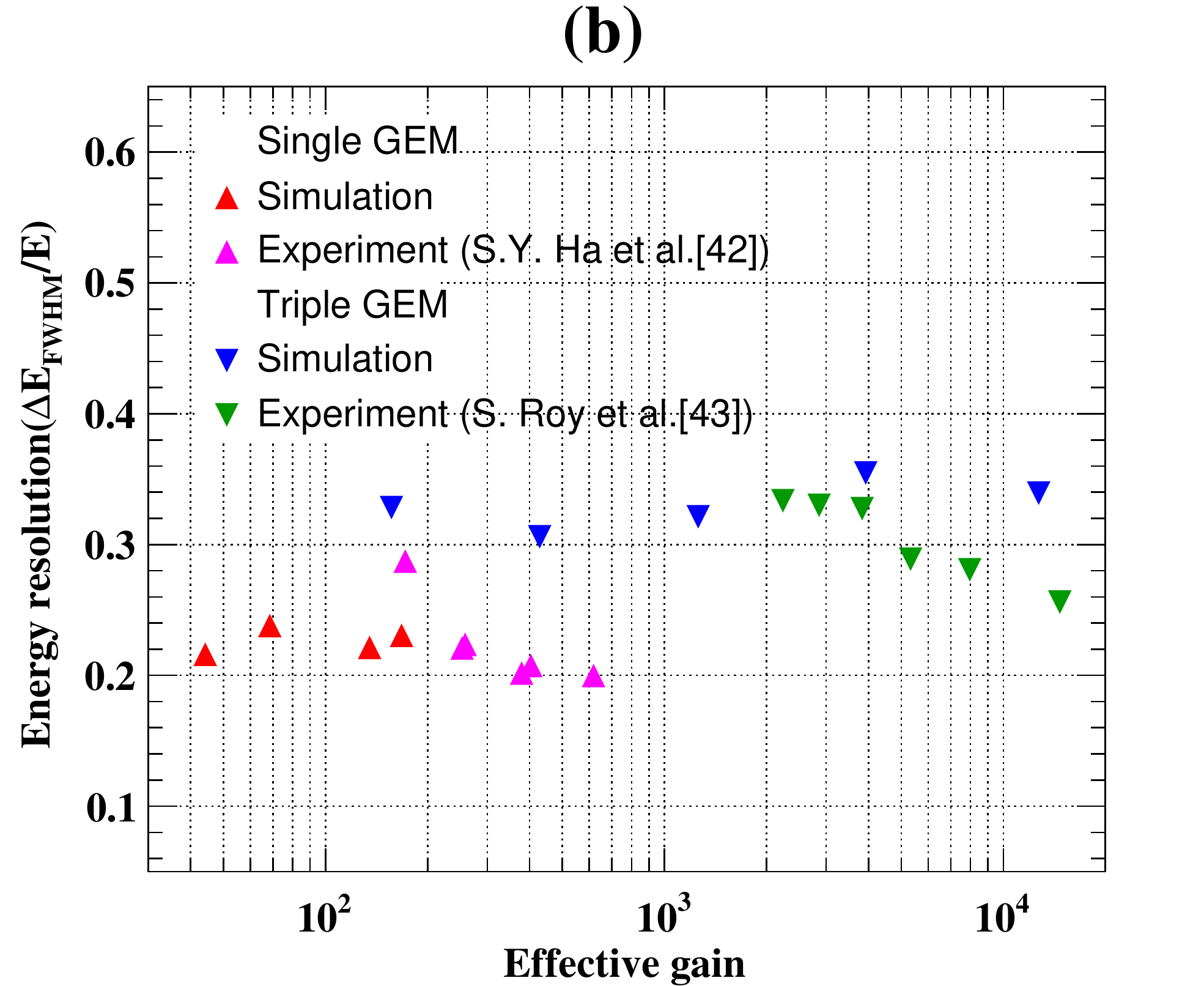}
\caption{Comparison of numerical estimates of energy resolution with experimental measurements by \cite{S.Y,Roy:2018ept}, as a function of (a) different applied voltages to GEM foils and (b) effective gain for single and triple GEM detectors.}
\label{fig:GainResolution}
\end{figure}

In a real experimental scenario, all possible cases of primary seed cluster variations and number fluctuations of primaries are likely to occur.
However, in the present model, the probability of occuring only these five different cases of cluster spread and the number variations of primaries have been taken into account. 
Figure \ref{fig:GainResolution} (a) shows the numerical estimates of energy resolution values as a function of applied voltages ($\Delta{V}_{GEM}$) to the GEM foil(s) of single and triple GEM detectors and compared with experimentally observed values reported in \cite{S.Y,Roy:2018ept}.
Figure \ref{fig:GainResolution} (b) shows the numerical estimates of energy resolution values as a function of effective gain.
It may be noted here that, the experiments \cite{S.Y} and \cite{Roy:2018ept} used for comparison in figures \ref{fig:GainResolution} (a) and (b) have geometrical parameters and experimental conditions different from those used in the simulation. 
Despite the differences in experimental conditions and simplifying assumptions in the numerical model, a reasonable agreement between simulation and experiment has been achieved.  
The simulated values of energy resolution for a single GEM is found to vary from 21\% to 24\%, whereas for a triple GEM detector it varies from 30\% to 35\%. 

\subsection{Primary ionization}
\label{subsec:Primary}
From the brief study in the preceding subsection, it is concluded that the proposed model is capable of estimating statistically fluctuating properties of GEM-based detectors to a reasonable accuracy.
In the remaining part of section \ref{sec:Results}, numerical estimate of discharge probability will be presented using an $^{241}$Am radiation source which emits 5.6 MeV alpha particles into the Ar$-$CO$_{2}$ gas mixture in volume proportions 70$-$30. 
The simulation has been carried out for single and triple GEM structures under different applied voltages as mentioned in table \ref{tab:table1}. 
The choice of geometrical parameters of the detectors and the radiation source used follows the experiment described in ~\cite{Bachmann:2000az}. 

\begin{figure}[htbp]
\centering
\includegraphics[width=0.49\linewidth]{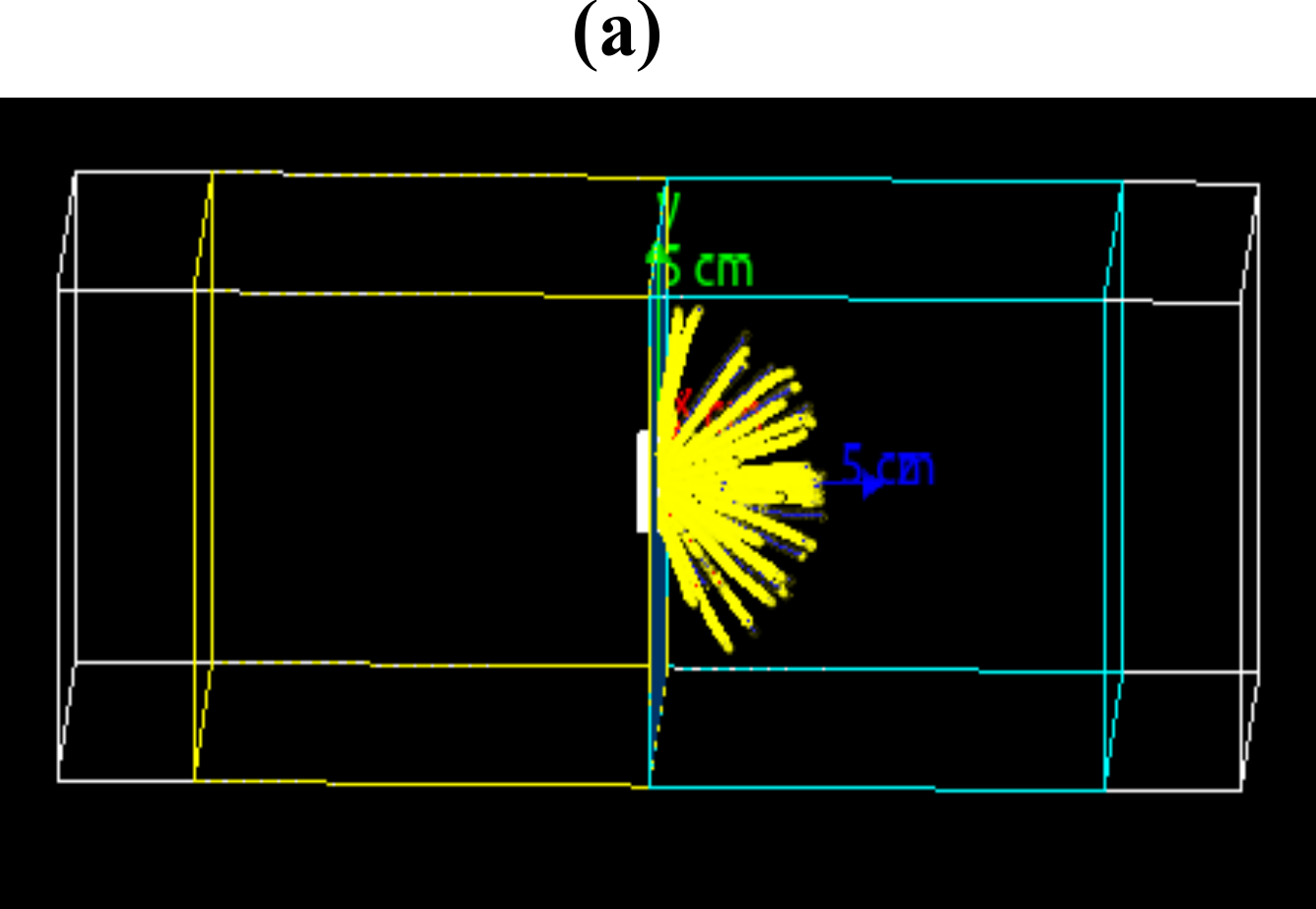}
\includegraphics[width=0.49\linewidth]{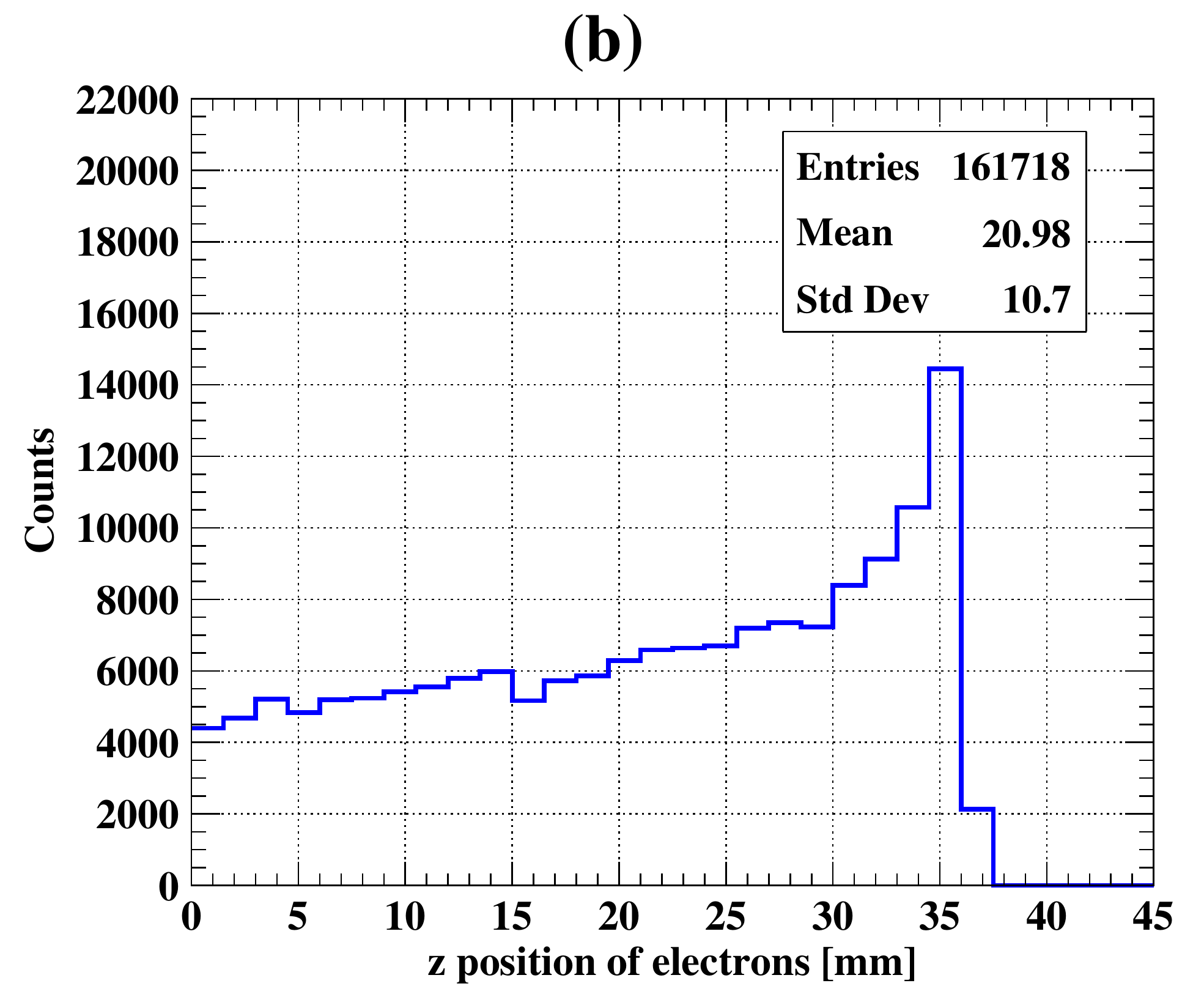}
\caption{(a) Geant4 display of the collimated alpha tracks with an angle of $\pm$ 30$^{\circ}$ in Ar$-$CO$_{2}$ (70-30) gas volume. (b) z position of primary electrons in an event}
\label{fig:G4GeomPrimary}
\end{figure}

A total of 10000 events have been simulated using the Geant4 model described in section \ref{subsec:EventGen}.
Tracks resulting from several alpha events, as simulated by the model, have been shown in figure \ref{fig:G4GeomPrimary}(a). 
For each event, an alpha particle deposits more energy towards the end of the trajectory in the gas volume.
As a result, the number density of primaries increases near the Bragg peak.
The z position of primaries for an event is shown in figure \ref{fig:G4GeomPrimary}(b).
It may be noted that the number of primaries generated in this particular event is nearly $1.6 \times 10^{5}$ which is close to the expected value obtained by dividing the energy of the alpha beam by the effective ionization potential of the gas mixture \cite{Gasik:2017uia}.
The range of an alpha particle along the z direction is nearly 4 cm in the gas volume as shown in figure \ref{fig:G4GeomPrimary}(b) which is also reasonably close to the value mentioned in \cite{Gasik:2017uia}.
The differences among the values reported in \cite{Gasik:2017uia} and the present work are possibly due to differences in simulation details, including the absence / presence of collimation, mylar sheet etc.
\subsection{Method of estimating discharge probability}
\label{subsec:Method}
\begin{figure}[htbp]
\centering
\includegraphics[width=0.49\linewidth]{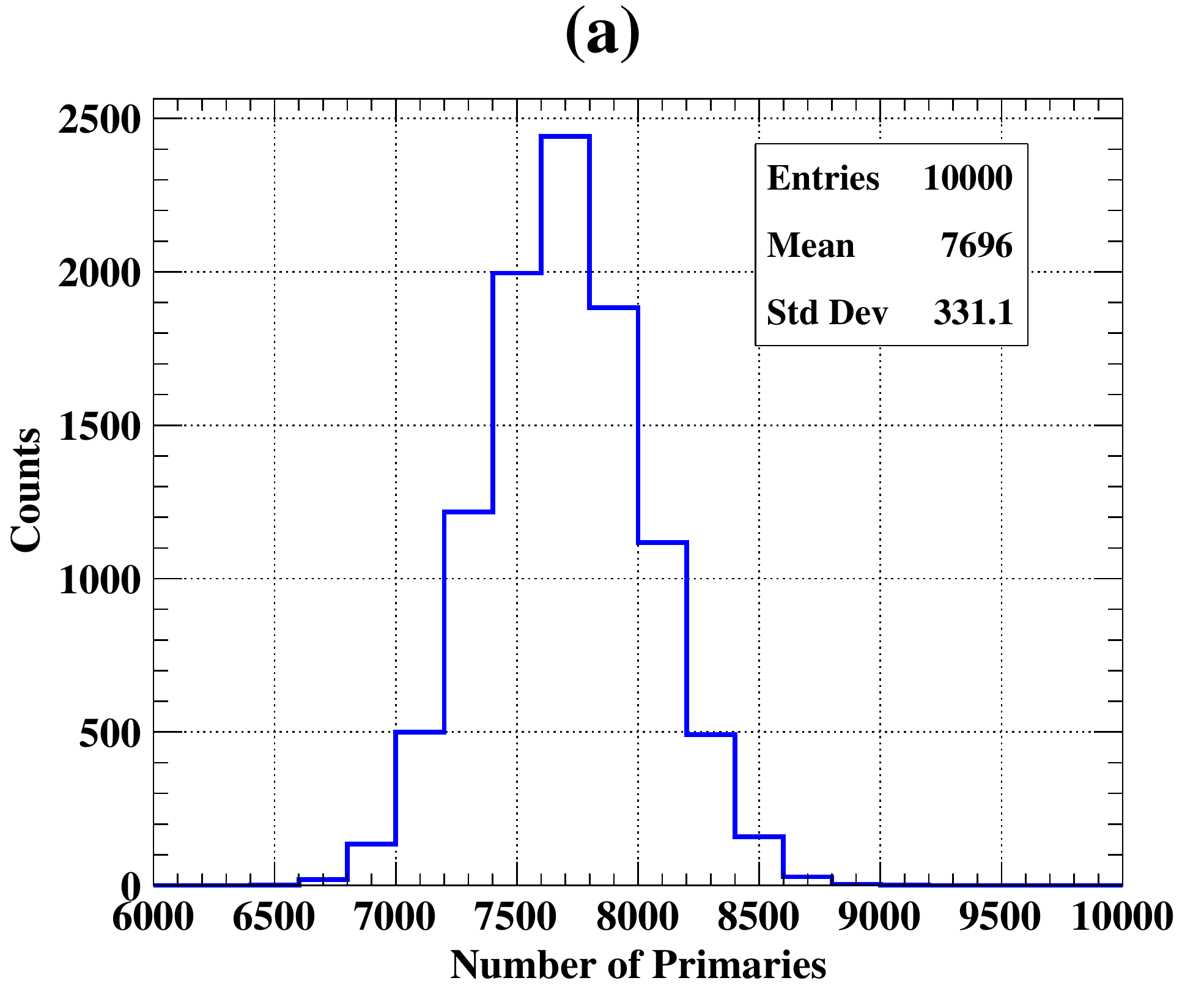}
\includegraphics[width=0.49\linewidth]{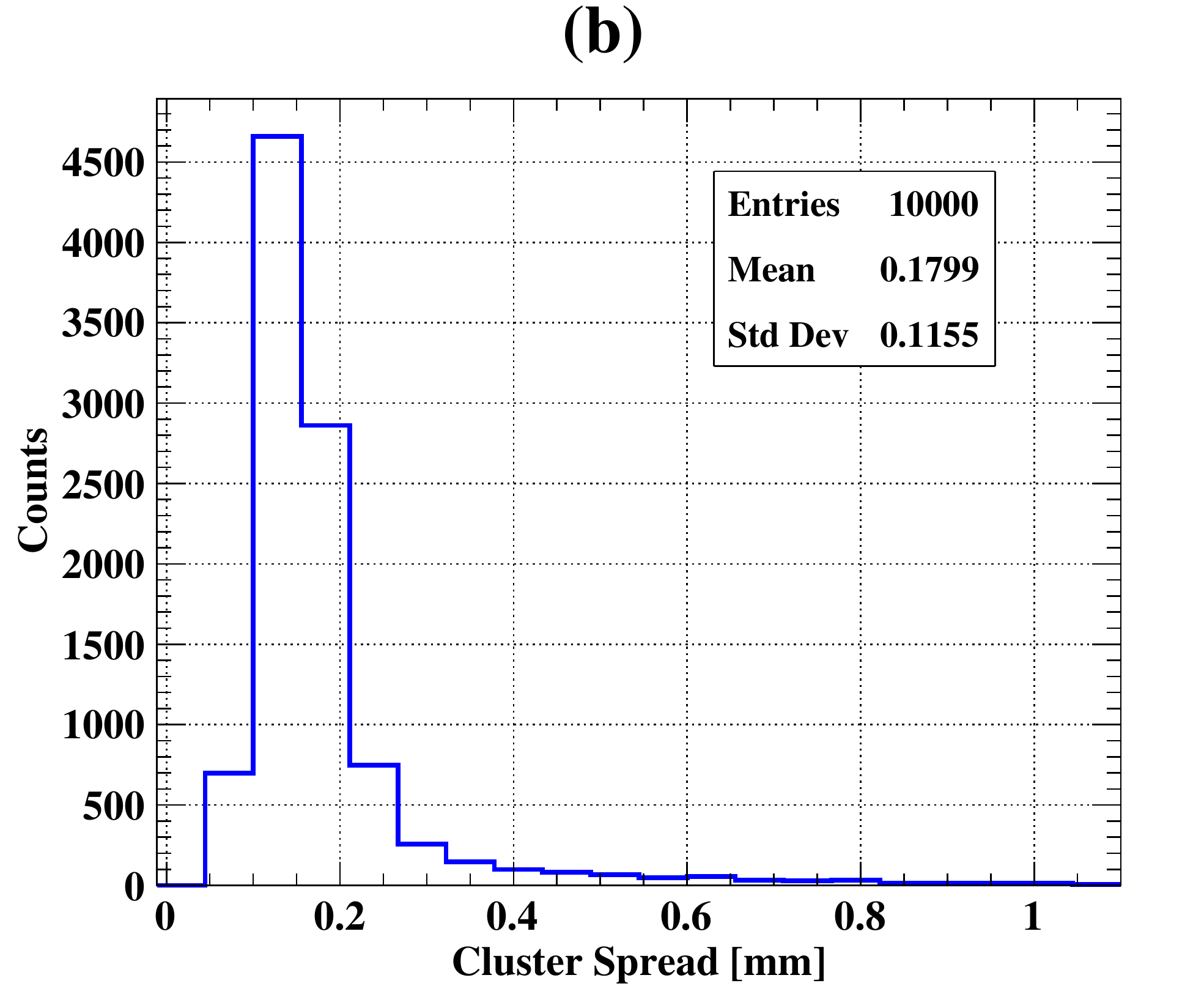}
\includegraphics[width=0.49\linewidth]{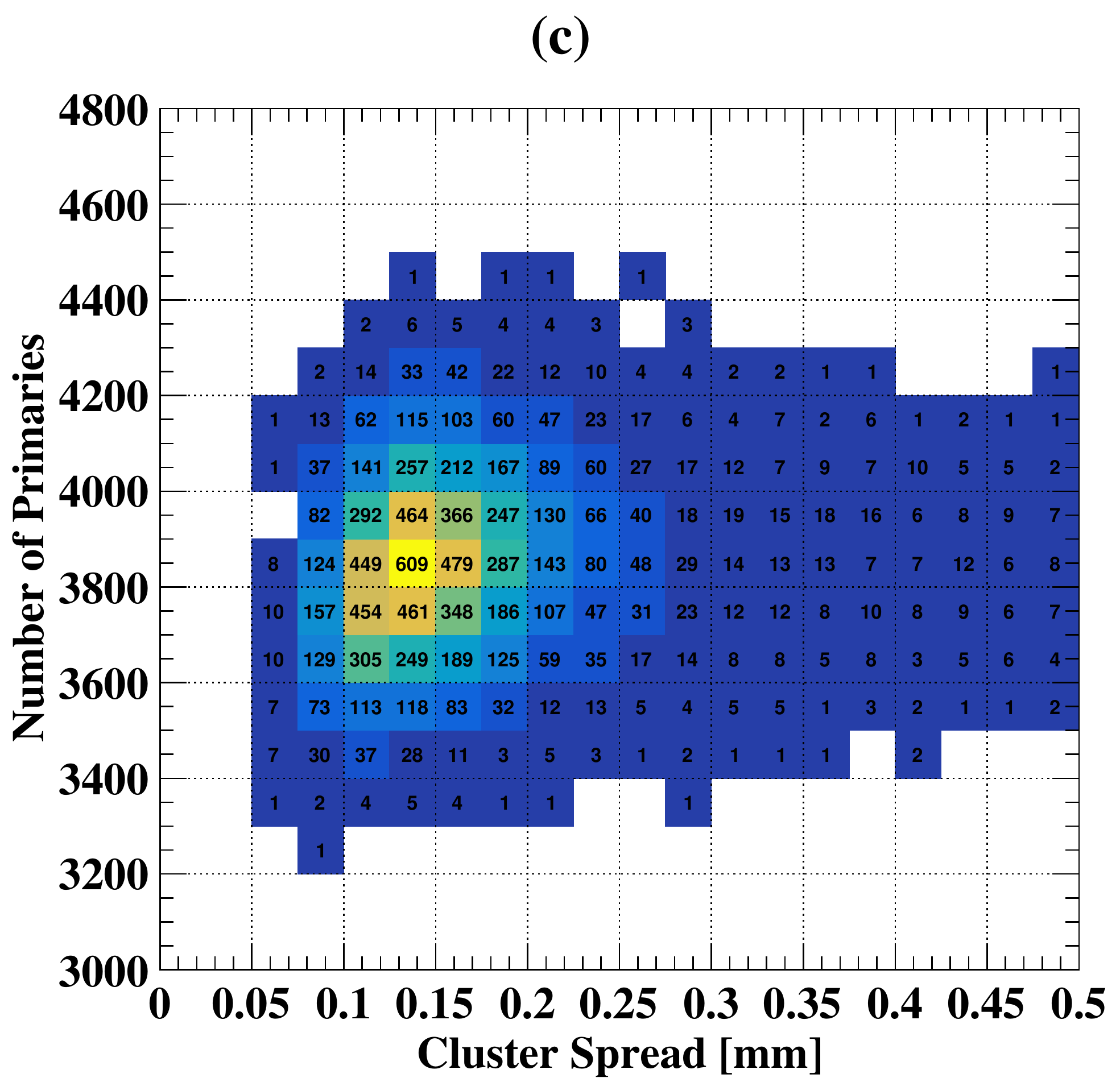}
\includegraphics[width=0.49\linewidth]{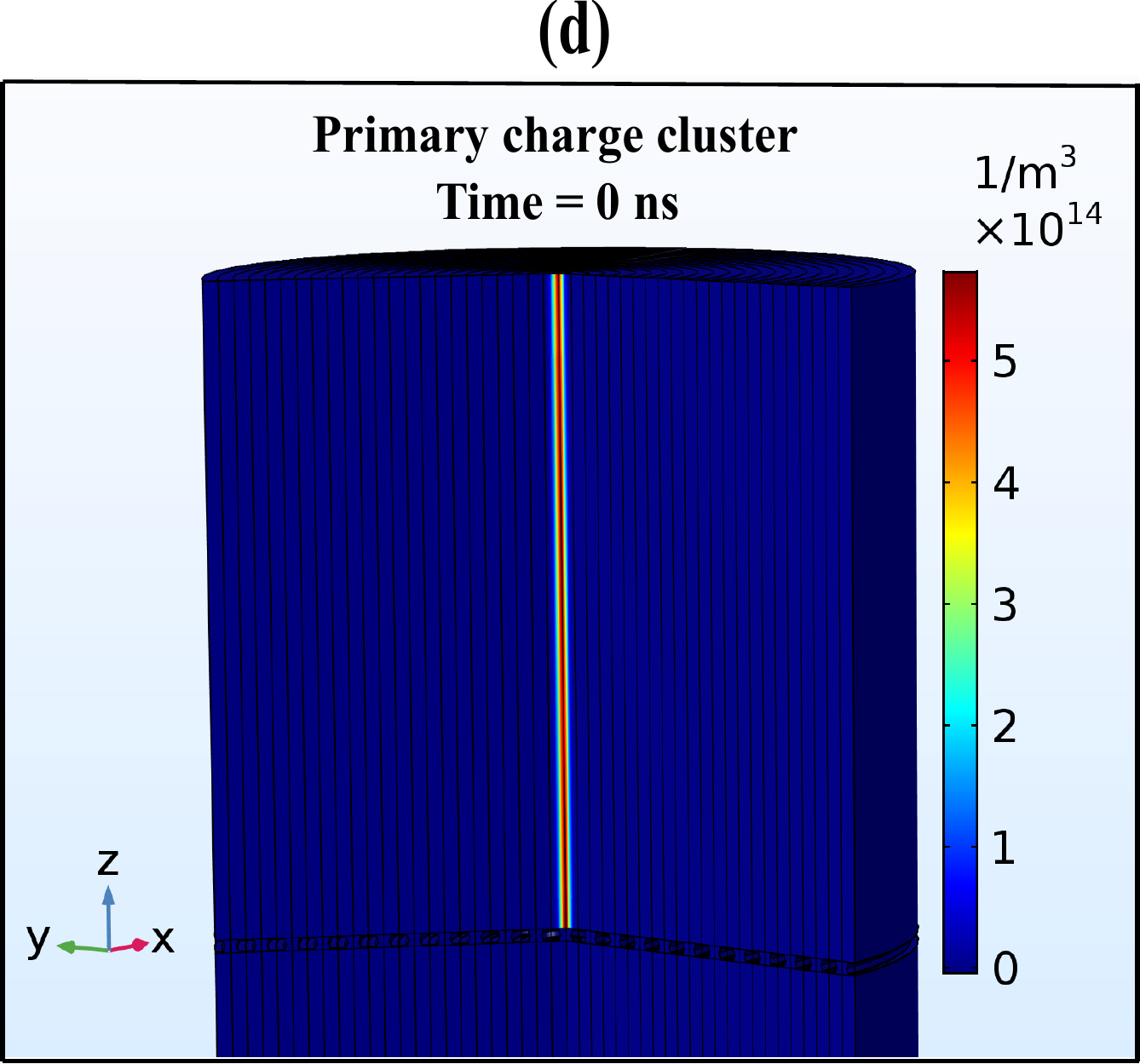}
\caption{(a) Distribution of number of primaries within 3 mm gas gap for 10000 events. (b) Cluster spread distribution in 3mm gas gap for 10000 events. (c) 2D histogram of number of primaries and cluster spread in 3mm gas gap. (d) Primary charge cluster for $^{241}$Am alpha source in the 3 mm drift gap of a single GEM detector at t = 0 ns}
\label{fig:AlphaPrimaries}
\end{figure}

The hydrodynamic simulation has been initiated utilizing information related to the number fluctuation and spatial distribution  of the primaries, as obtained from the preceeding section \ref{subsec:Primary}.
According to ~\cite{Bachmann:2000az}, the alpha source was placed external to the drift electrode, which is 3 mm above the first GEM foil.
Thus, it is assumed that the number of primaries generated in the first 3 mm (z direction) of the simulated gas volume will correspond to the drift gap in the experiment and will contribute to avalanches, or discharges observed in the experiment.
Figure \ref{fig:AlphaPrimaries}(a) shows the distribution of number of primaries in each event within the 3 mm gas gap. 
The cluster spread for each event has been obtained by subtracting the maximum and the minimum radial positions of the primary electons and is presented in figure \ref{fig:AlphaPrimaries}(b). 
It may be noted that the primary electrons are uniformly distributed over the 3 mm gas gap along the z direction as shown in figure \ref{fig:G4GeomPrimary}(b). 
For the axisymmetric model considered in this work, the number of electrons being collected by the central hole determines the evolution of the subsequent transport and amplification of the charged species.
Because of different transport processes in the drift region, number of electrons are lost in the copper surfaces of the GEM foil and significant percentage of electrons enter into off-centre holes.
According to charge sharing data computed following ~\cite{Roy:2020gwl}, around 50\% of primary electrons are found to enter into the central hole. 
Thus, for each event, the number of primary electrons contributing to the gas amplification process in the central hole has been considered to be 50\% of the number indicated in figure \ref{fig:AlphaPrimaries}(a).
Figure \ref{fig:AlphaPrimaries}(c) presents a 2D histogram that relates the number of primary electrons in an event with the cluster spread.
The initial seed cluster has been represented by a uniform distribution along z direction and gaussian distribution along the radial direction of the 2D axisymmetric model.
The charged fluid computed at the initial moment of the simulation at t = 0 ns has been shown in the figure \ref{fig:AlphaPrimaries}(d).
The volume integral of the primary cluster gives the number of primaries present in the simulation volume at the initiation of the hydrodynamic simulation.

The evolution of the charged species and possibility of achieving either an avalanche or discharge depend on the number of primaries in the drift volume and corresponding spatial distribution, among other simulation parameters, such as potential configuration of the detector etc. 
Thus, for a given applied voltage to the GEM foil(s), the evolution of the charged fluid has been simulated for several possible combinations of primary electron numbers and cluster spread. 
Tables \ref{tab:table4} and \ref{tab:table5} show various combinations of number of primaries and cluster spreads that have been utilized in the simulation for possibility of discharges, with different applied voltages to the GEM foil in single and triple GEM axisymmetric model.
Following figure \ref{fig:AlphaPrimaries}(a) and the discussion in the preceding paragraph, the range of number of primary electrons has been considered to be 3200 to 4500, in steps of 100.
Similarly, following figure \ref{fig:AlphaPrimaries}(b), the cluster spread for the simulations presented in the tables ranges between 0.05mm to 1.1mm with a binning of 0.025mm.
For a certain voltage, at each value of cluster spread in the range, simulations have been performed for all possible numbers of primaries, mentioned as \textbf{All} in the table.
Out of all possible combinations, some are found for which the discharge occurs, indicated by \textbf{Yes}.
The remaining number of primaries (\textbf{Rest}) does not lead to any discharge, referred to as \textbf{No} in the table.
According to the present estimations, for a given voltage, a combination of higher number of primaries with less cluster spread is found to be more effective in producing discharges.
Similar observations have also been reported in [15].
However, there are voltage thresholds below which no allowed combination of number of primaries and cluster spread lead to a discharge.
Similar observations have also been reported in ~\cite{Bachmann:2000az}.

\subsection{Discharge probability estimates}
\label{subsec:Discharge}
\begin{figure}[htbp]
\centering
\includegraphics[width=0.49\linewidth]{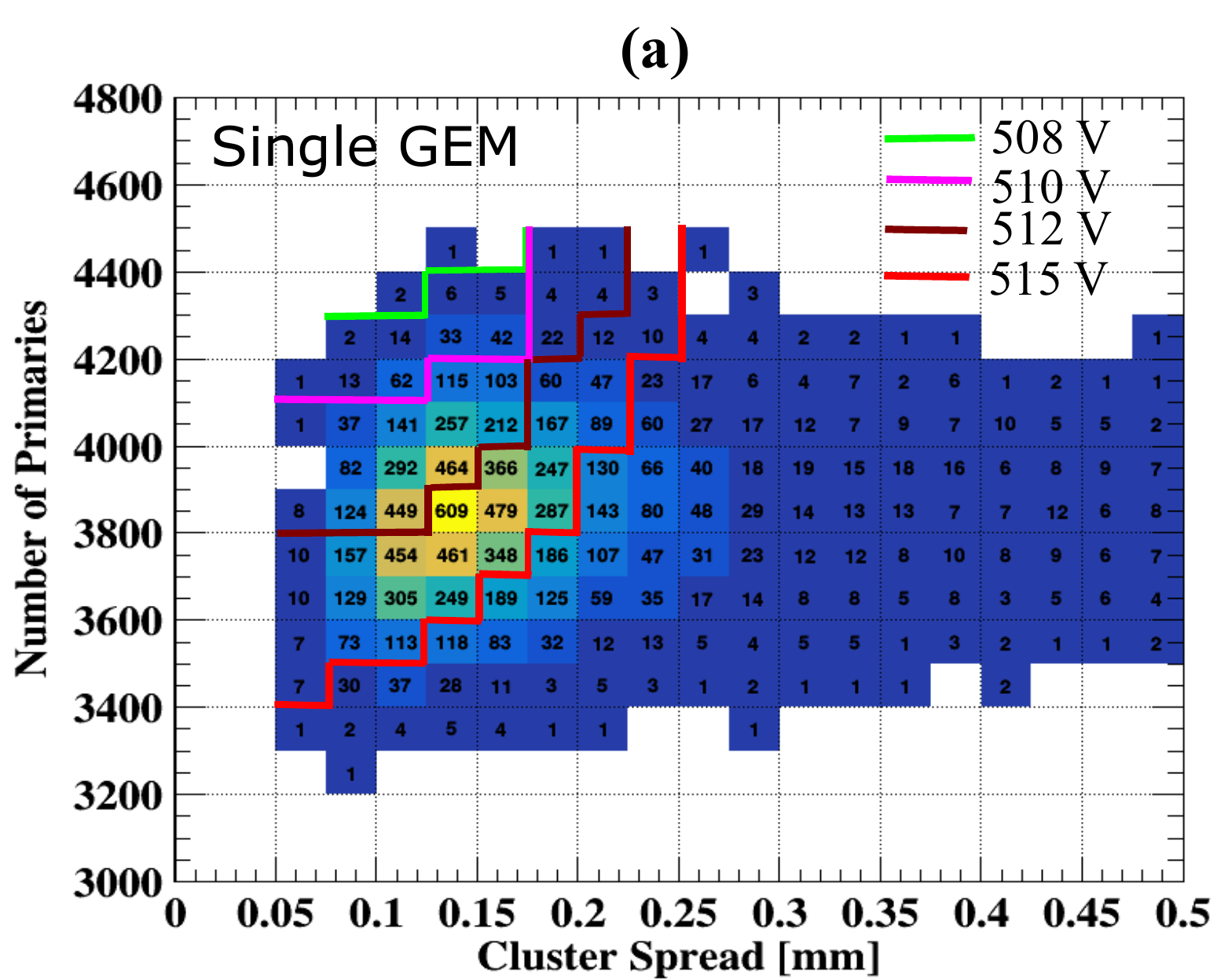}
\includegraphics[width=0.49\linewidth]{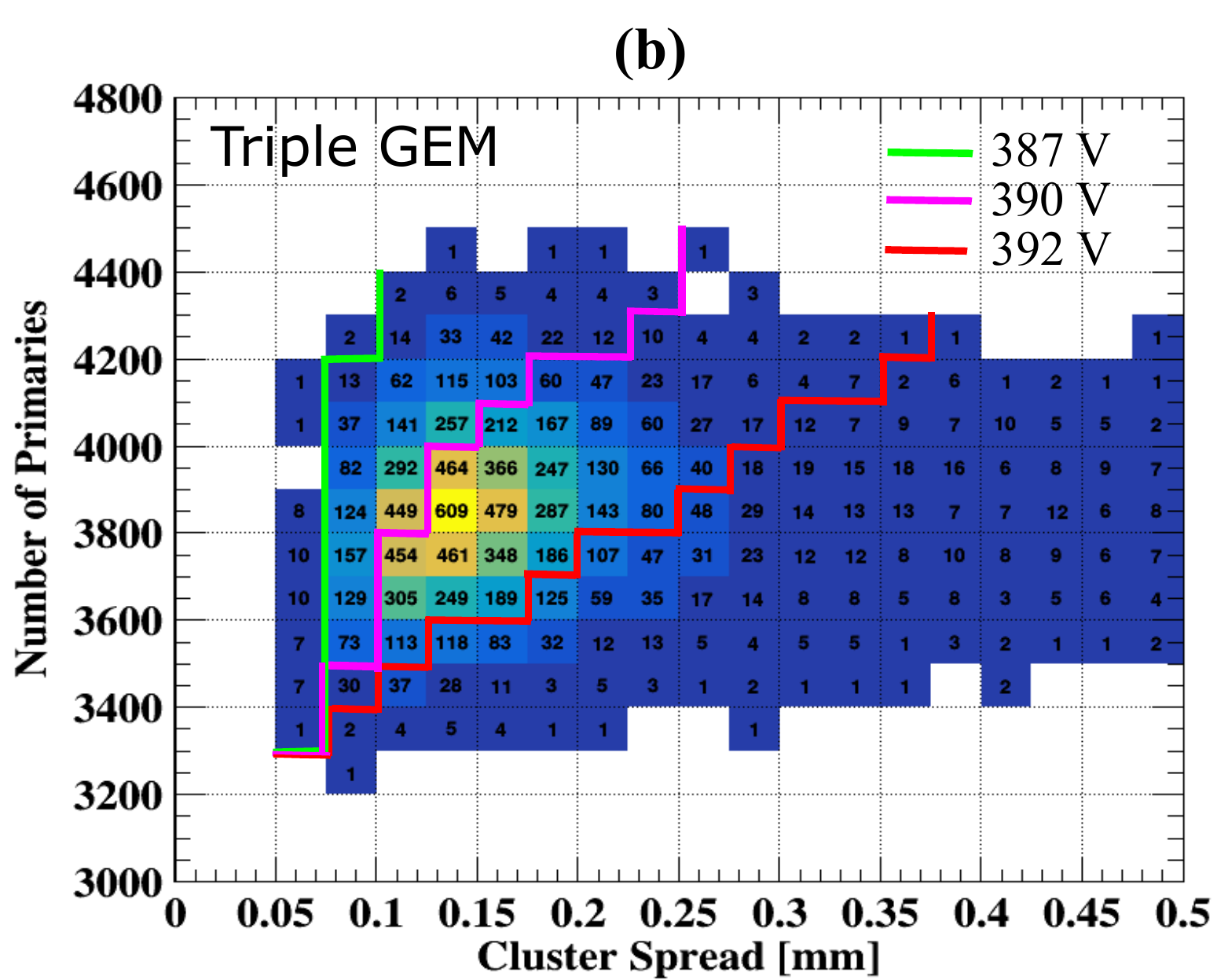}
\caption{Curves are drawn over the 2D histogram of cluster spread and number of primaries using different colors showing different voltages applied to the GEM foil(s) of (a) single GEM and (b) triple GEM for identifying the discharge events. The discharge occurs for the number of events present above the curves indicated in different color bars. The events present below each curve do not give discharge.}.    
\label{fig:Discharge_curves}
\end{figure}

As described in ~\cite{Rout:2020bjz}, discharges are observed in the central hole due to the accumulation of ionic space charges
in which the flow of charged fluids from GEM anode to GEM cathode is associated with strong distortion of appied electric field.
The occurrence of such discharges happen in a small fraction of all possible events which include all the possible alpha tracks, as computed by the Geant4 model described earlier in sections \ref{subsec:EventGen} and \ref{subsec:Primary}.
Figures \ref{fig:Discharge_curves} (a) and (b) show different number of cases of cluster spread and number of primaries, for which several executions of the simulations have been performed to identify the discharge events.
In these two figures, curves are drawn with different color bars showing different voltages applied to the GEM foil(s) of single and triple GEM, over a 2D histogram of cluster spread and number of primaries.
Events with greater number of primaries (above the boundary line) and less cluster spread (left of the boundary line) than those indicated by the coloured curves lead to discharges.  
Finally, the discharge probability in the present model has been calculated by following a similar definition used in the numerical and experimental studies by \cite{Gasik:2017uia}.
For a certain voltage configuration, the number of possible discharge events have been estimated by the hydrodynamic model.
From Geant4 we have estimated the total number of possible alpha events.
The ratio of the discharge events from the hydrodynamic model to the total number of possible alpha events from Geant4 gives rise to the discharge probability.

\begin{figure}[htbp]
\centering
\includegraphics[width=0.49\linewidth]{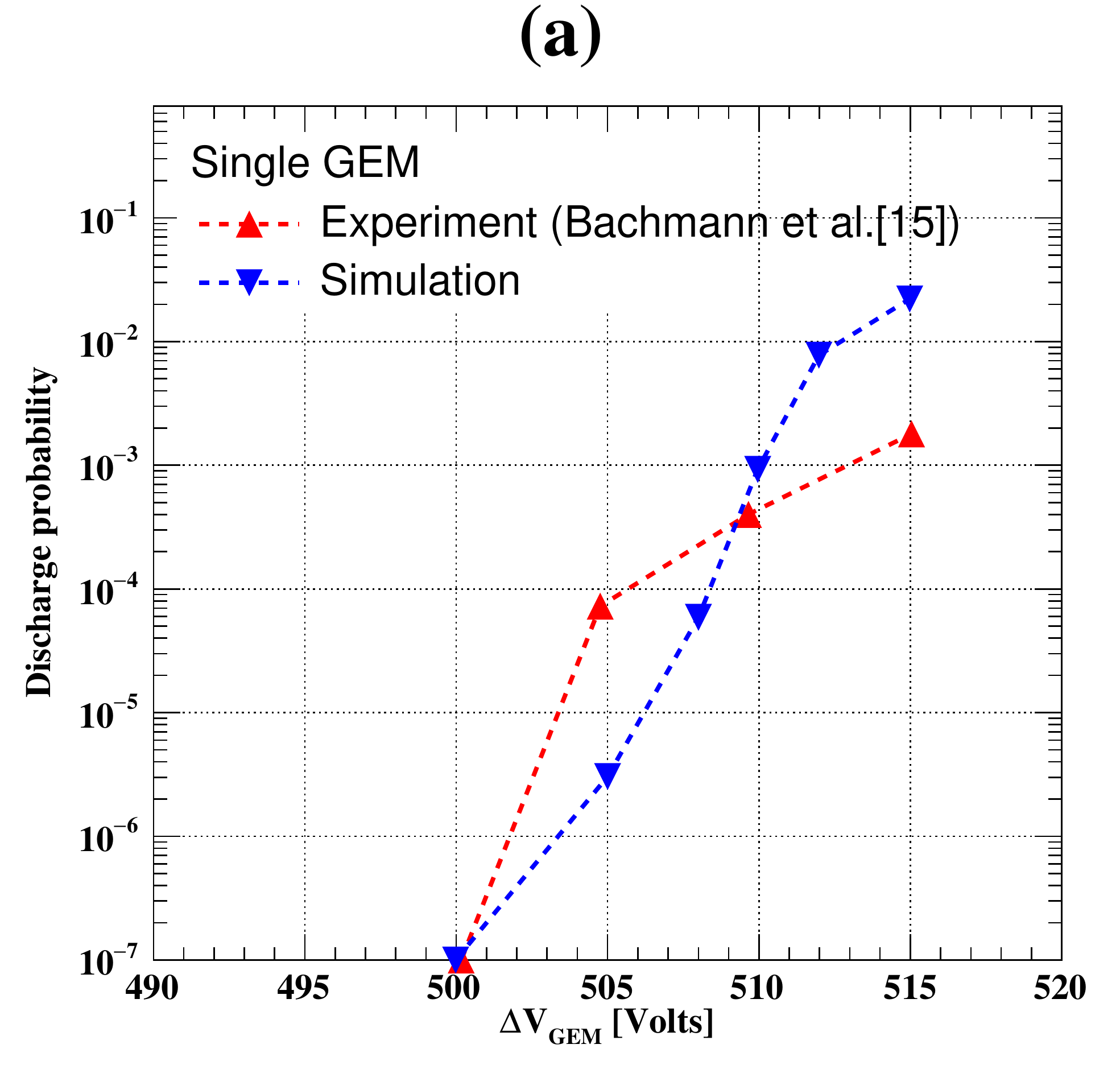}
\includegraphics[width=0.49\linewidth]{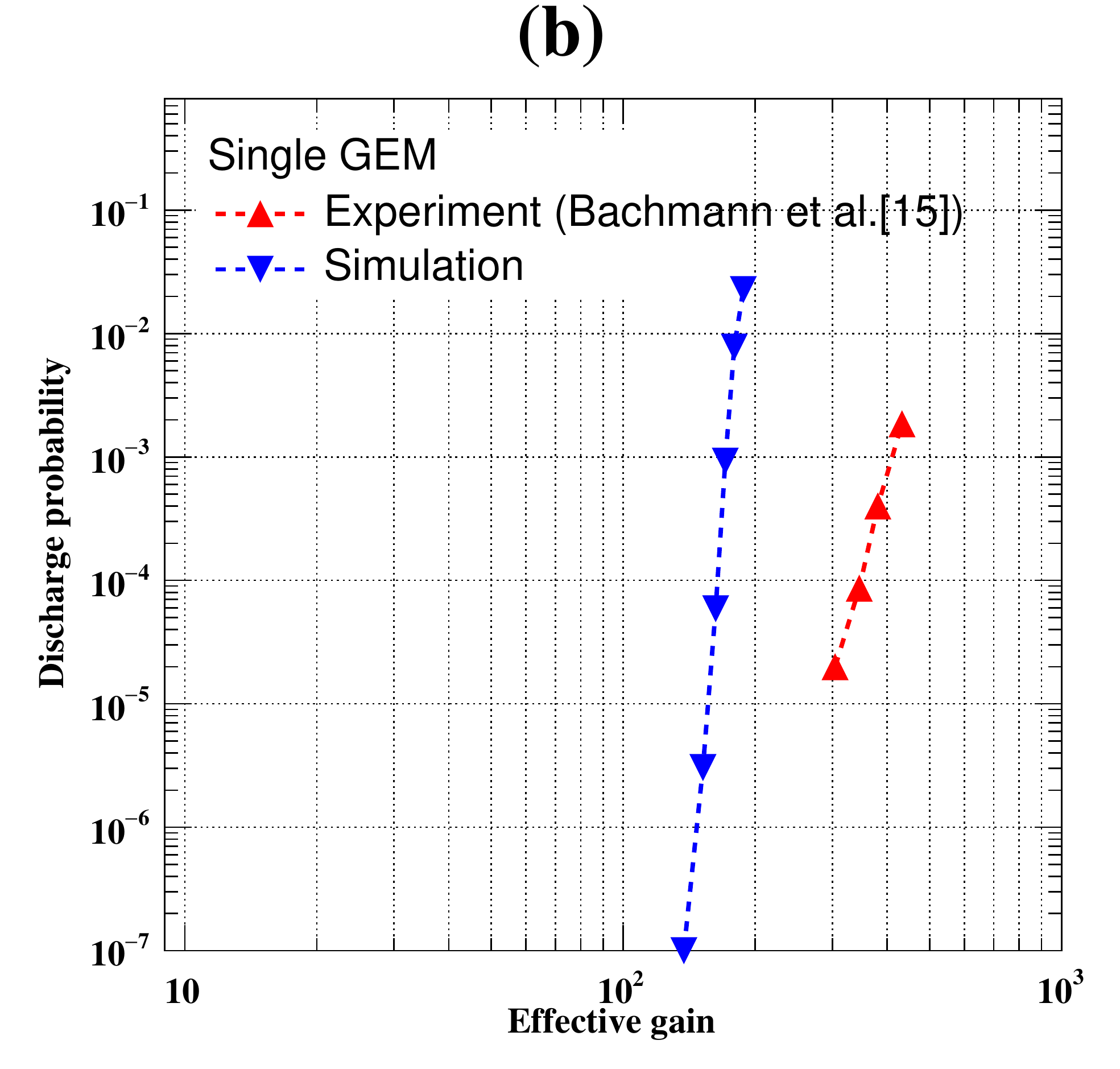}
\includegraphics[width=0.49\linewidth]{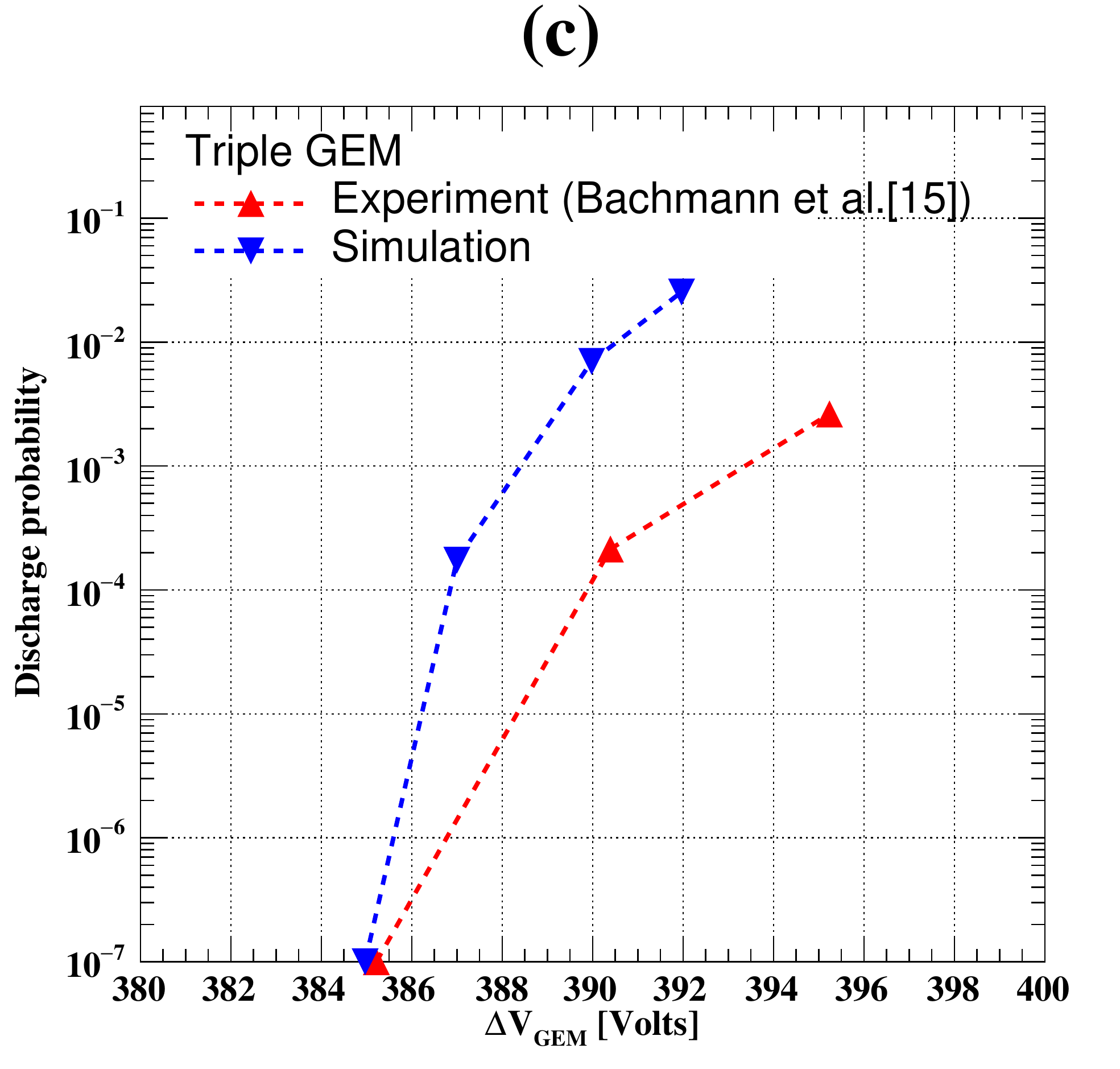}
\includegraphics[width=0.49\linewidth]{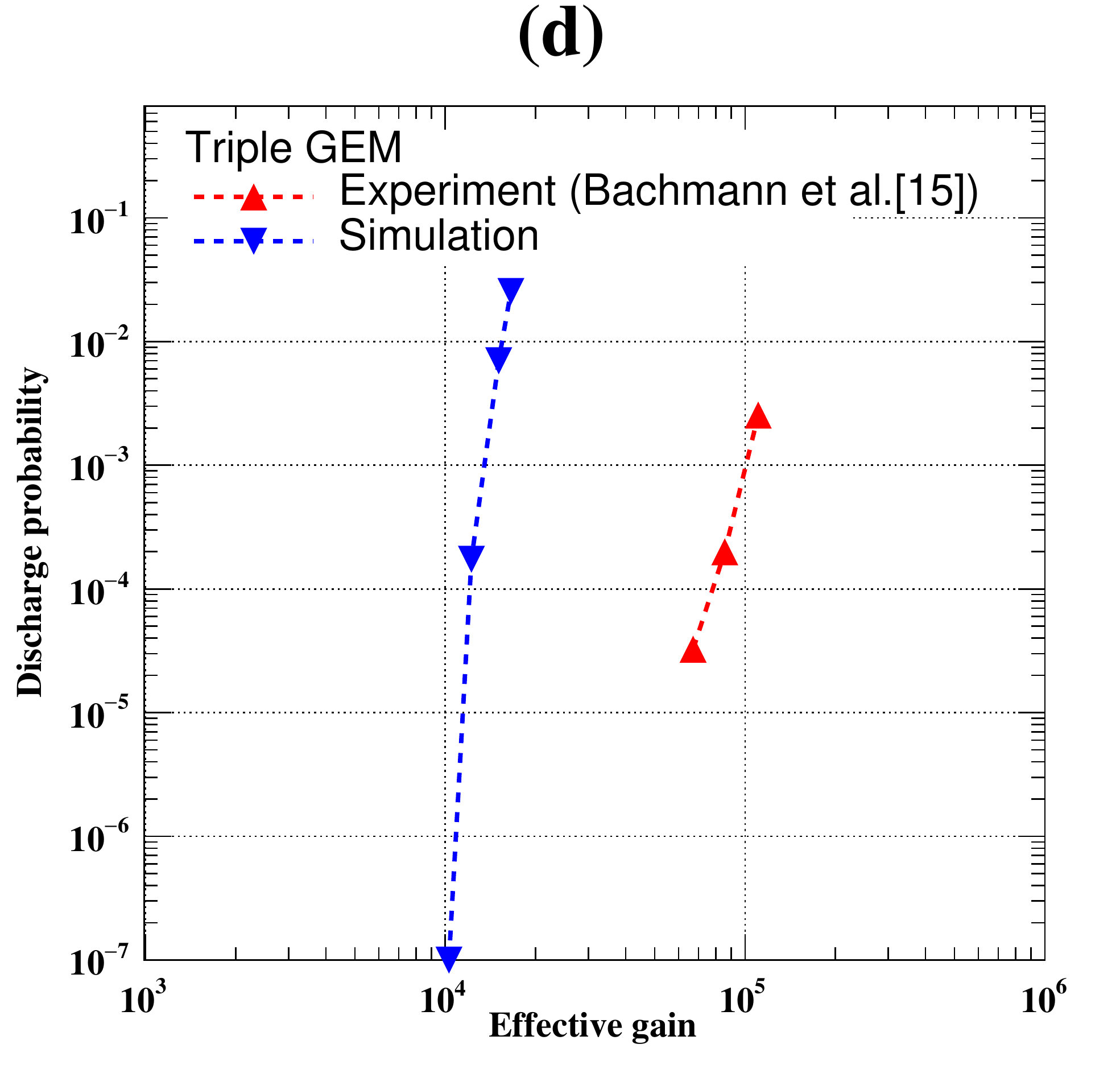}
\includegraphics[width=0.49\linewidth]{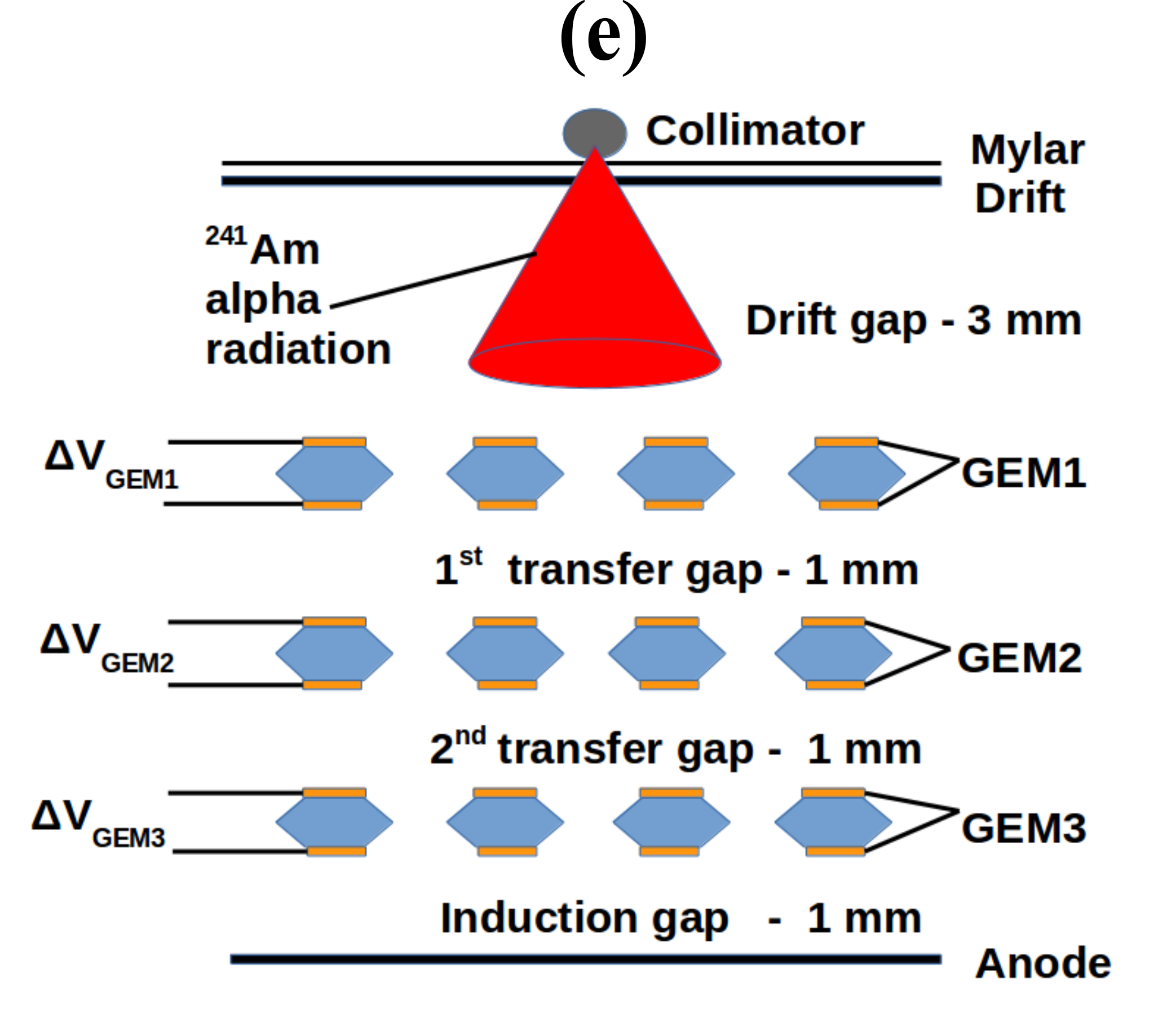}
\includegraphics[width=0.49\linewidth]{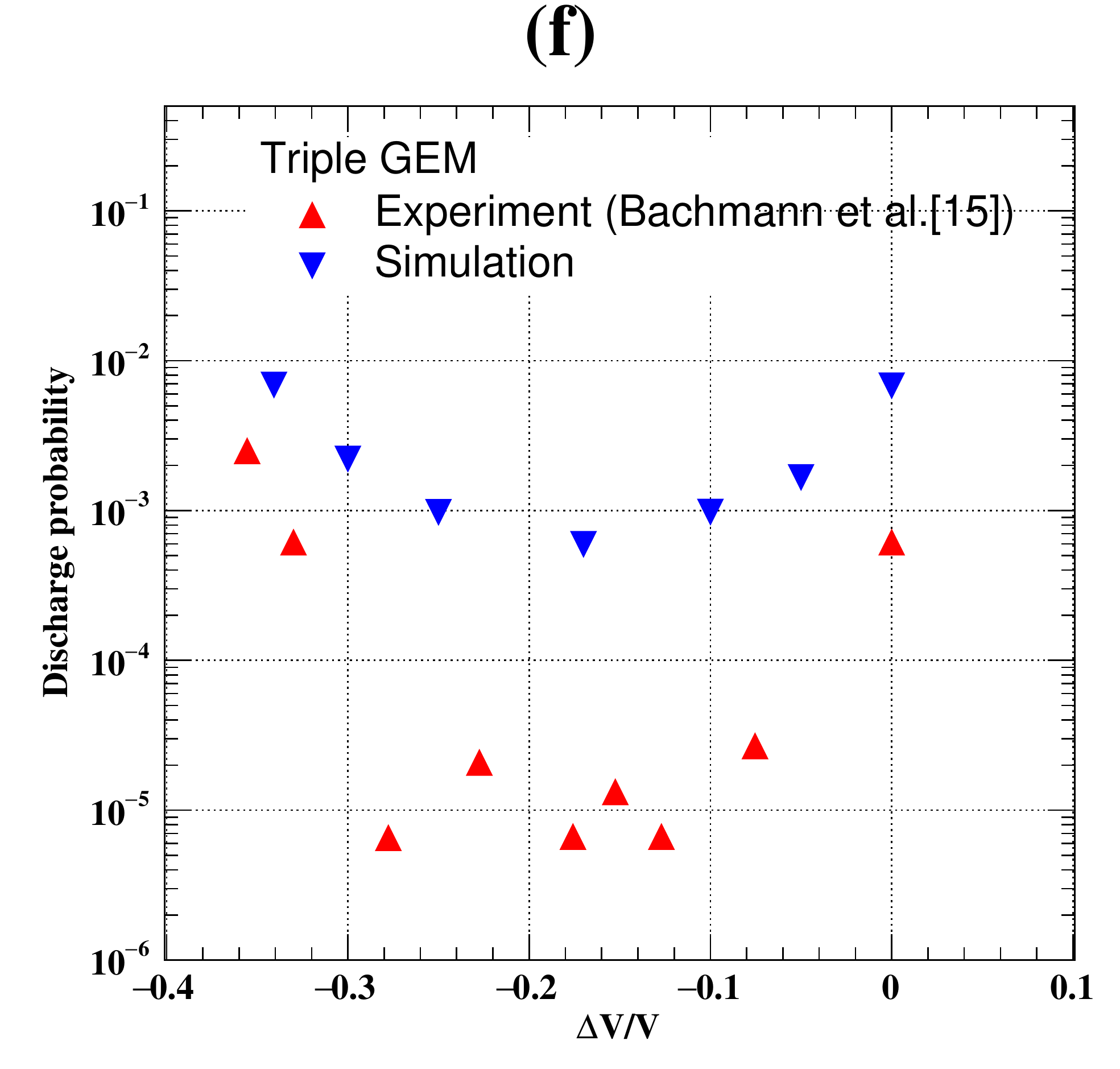}
\caption{Discharge probability estimates in single GEM (a) and (b) obtained from simulation and compared with experimental measurements reproduced from ~\cite{Bachmann:2000az}. Discharge probability estimates in triple GEM with equal voltages applied to three GEM foils in (c) and effective gain in (d). (e) Schematic diagram of triple GEM configuration. (f) Discharge probability estimates with asymmetric distribution of voltages obtained from simulation and  compared with experimental measurements reproduced from ~\cite{Bachmann:2000az}.}
\label{fig:DischargeProbability}
\end{figure}

The resulting discharge probability values have been compared to experimental results presented in ~\cite{Bachmann:2000az}.
Figures \ref{fig:DischargeProbability}(a) and (b) show the discharge probability estimates for single GEM detector with different applied voltages and effective gain and compared with the experimental measurements reproduced from ~\cite{Bachmann:2000az}. 
Similarly, figures \ref{fig:DischargeProbability}(c) and (d) show the discharge probability estimates for triple GEM detector with applied voltages and effective gain and compared with the experimental measurements reproduced from ~\cite{Bachmann:2000az}.
From these figures, it is observed that the discharge probability values in the voltage ranges considered follow the experimental trend for both single and triple GEM detectors. 
However, the values obtained from the simulation do not exactly match the experimental values, the simulated ones having over-estimated the possibility of discharges.

\begin{table}[h!]
  \begin{center}
    \begin{tabular}{|c|c|c|c|}
    \hline
      \textbf{Applied Voltages} & \textbf{Cluster Spread} & \textbf{Number of Primaries} & \textbf{Discharge}\\
       $\Delta{V}_{\rm GEM}$ (Volts) & (mm) & Range  & Yes (No)\\
      \hline
      500  & 0.05 - 1.1 & All & No \\
      \hline
      505 & 0.05 - 1.1 & All & No \\
      \hline
      508 & 0.050 - 0.085  & All & No \\
          & 0.085 - 0.125  & 4300 - 4400 (Rest) & Yes (No) \\
          & 0.125 - 0.175  & 4400 (Rest) & Yes (No) \\
          & 0.175 - 1.1    & All & No \\
      \hline
      510 & 0.050 - 0.125 & 4100 - 4400 (Rest) & Yes (No) \\
          & 0.125 - 0.175 & 4200 - 4400 (Rest) & Yes (No) \\
          & 0.175 - 1.1 & All  & No \\
      \hline
      512 & 0.050 - 0.125 & 3800 - 4400 (Rest) & Yes (No) \\
          & 0.125 - 0.150 & 3900 - 4400 (Rest) & Yes (No) \\
          & 0.150 - 0.175 & 4000 - 4400 (Rest) & Yes (No) \\
          & 0.175 - 0.200 & 4200 - 4400 (Rest) & Yes (No) \\
          & 0.200 - 0.225 & 4300 - 4400 (Rest) & Yes (No) \\
          & 0.225 - 1.1   & All & No \\
       \hline
       515 & 0.050 - 0.075 & 3400 - 4400 (Rest) & Yes (No) \\
           & 0.075 - 0.125 & 3500 - 4400 (Rest) & Yes (No) \\
           & 0.125 - 0.150 & 3600 - 4400 (Rest) & Yes (No) \\
           & 0.150 - 0.175 & 3700 - 4400 (Rest) & Yes (No) \\
           & 0.175 - 0.200 & 3800 - 4400 (Rest) & Yes (No) \\
           & 0.200 - 0.225 & 4000 - 4400 (Rest) & Yes (No) \\
           & 0.225 - 0.250 & 4200 - 4400 (Rest) & Yes (No) \\
           & 0.250 - 1.1   & All & No \\
       \hline        
    \end{tabular}
    \caption{Different combination of number of primaries and cluster spread values for getting possible discharge events in single GEM with different voltages applied to the GEM foil. For a given range of cluster spread values, out of all primaries ("All"), a range of primaries gives rise to discharges and is mentioned as "Yes". The rest of the primaries "Rest" do not give rise to discharges and is indicated as "No".}
    \label{tab:table4}
  \end{center}
\end{table}
\begin{table}[h!]
  \begin{center}
    \begin{tabular}{|c|c|c|c|}
    \hline
      \textbf{Applied Voltages} & \textbf{Cluster Spread} & \textbf{Number of Primaries} & \textbf{Discharge}\\
       $\Delta{V}_{\rm GEM}$ (Volts) & (mm) & Range & Yes (No)\\
      \hline
      385  & 0.05 - 1.1 & All & No  \\
      \hline
      387 & 0.05  - 0.075  & All & Yes \\
          & 0.075 - 0.100 & 4200 - 4400 (Rest) & Yes (No) \\
          & 0.100 - 1.1 & All & No  \\
      \hline
      390 & 0.05  - 0.075 & All & Yes \\
          & 0.075 - 0.100 & 3500 - 4400 (Rest) & Yes (No)  \\
          & 0.100 - 0.125 & 3800 - 4400 (Rest) & Yes (No) \\
          & 0.125 - 0.150 & 4000 - 4400 (Rest) & Yes (No) \\
          & 0.150 - 0.175 & 4100 - 4400 (Rest) & Yes (No) \\
          & 0.175 - 0.225 & 4200 - 4400 (Rest) & Yes (No) \\
          & 0.225 - 0.250 & 4300 - 4400 (Rest) & Yes (No) \\
          & 0.250 - 1.1   & All & No \\
     \hline
      392 & 0.05  - 0.075 & All & Yes \\
          & 0.075 - 0.100 & 3300 - 4400 (Rest) & Yes (No) \\
          & 0.100 - 0.125 & 3400 - 4400 (Rest) & Yes (No) \\
          & 0.125 - 0.175 & 3600 - 4400 (Rest) & Yes (No) \\
          & 0.175 - 0.200 & 3700 - 4400 (Rest) & Yes (No) \\
          & 0.200 - 0.250 & 3800 - 4400 (Rest) & Yes (No) \\
          & 0.250 - 0.275 & 3900 - 4400 (Rest) & Yes (No) \\
          & 0.275 - 0.300 & 4000 - 4400 (Rest) & Yes (No) \\
          & 0.300 - 0.350 & 4100 - 4400 (Rest) & Yes (No) \\
          & 0.350 - 0.375 & 4200 - 4400 (Rest) & Yes (No) \\
          & 0.375 - 1.1   & All & No \\
      \hline 
    \end{tabular}
    \caption{Different combination of number of primaries and cluster spread values for achieving discharges in triple GEM. Equal voltages are applied to the three GEM foils in triple GEM. For a given range of cluster spread, out of all primaries ("All"), a range of primaries gives rise to discharges and is represented as "Yes". The rest of the primaries "Rest" do not give rise to discharges and is termed as "No".}
    \label{tab:table5}
    \end{center}
\end{table}

The simulation has also been performed by applying asymmetric voltages in three GEM foils of the triple GEM detector. 
A schematic diagram of triple GEM configuration utilized in the simulation for estimating the probability of discharge formation is shown in figure \ref{fig:DischargeProbability} (e).
The percentage difference in operating voltages [$\frac{\Delta V}{V} = \frac{\Delta{V}_{GEM3} - \Delta{V}_{GEM1}}{\Delta{V}_{GEM2}} $] has been utilized in the simulation and follows closely with the asymmetric configuration (+ 0 -) shown in figure 12 of \cite{Bachmann:2000az}.
The asymmetric configuration (+ 0 - ) utilized in simulation represents a fixed voltage of 390 V across GEM2 foil and higher (lower) voltages in GEM1 (GEM3) foils with an equal amount added/subtracted from $\Delta{V}_{GEM2}$ respectively. 
The discharge probability values obtained from simulation are compared with the experimentally observed values reproduced from the asymmetric configuration (+ 0 -) shown in figure 12 of \cite{Bachmann:2000az}. 
Figure \ref{fig:DischargeProbability} (f) shows the discharge probability as a function of the percentage difference in operating voltages applied to the three GEM foils of a triple GEM detector.
It may be noted that the general trend of the experimental data and numerical estimates matches qualitatively.
In general, the numerical estimate is higher than the experimentally observed values for discharge probability.

From the figure, fast increase in discharge probability is observed in experiment \cite{Bachmann:2000az} on either side of a minimum at $\frac{\Delta{V}}{V} \sim -0.17$, where the discharge probability is found to be less.
The same minimum is reproduced in the simulated results although the magnitude of the estimated discharge probability is almost two orders of magnitude more than experimental value.
It may be noted here that the simulation performed with equal voltages applied to three GEM foils in the voltage ranges, as mentioned in table \ref{tab:table1}, always leads to discharge formation in the third GEM foil.
However, discharge formation is observed in both GEM2 and GEM3 when using an asymmetric distribution of voltages in the three GEM foils.
Towards the negative values of ($\frac{\Delta{V}}{V}$) from 0, the applied voltage in GEM1 increased successively which enhanced the possibility for the formation of discharge in GEM2.   

\section{Conclusion}
\label{sec:Conclusion}
A hybrid fast numerical model has been proposed to estimate response of single and multiple GEM-based ionization detectors.
Information related to generation of primaries, collection efficiency and charge sharing have been computed using Monte-Carlo approach, while the transport and amplification of the charged species have been modelled using a hydrodynamic approach for an axisymmetric geometry.
Since the hydrodynamic model is deterministic in nature, sources of fluctuations have been introduced in terms of different possible combination of initial conditions.
Initial number of primaries, position and configuration of the seed cluster are found to have significant impact on the subsequent development of detector response.
Energy resolution of single and triple GEM detectors for 5.9 keV photons from the $^{55}$Fe radiation has been estimated.
Despite the simplifying assumptions and inherent limitations of the model, the estimated values are found to be reasonably close to experimental values reported elsewhere.
Attempts have been made, next, to estimate discharge probabilities in such detectors when exposed to highly ionizing alpha radiation.
The numerically estimated probabilities are found to follow the experimentally observed values in a qualitative manner.
Finally, the possibility of reducing discharge probability by applying asymmetric voltage distribution in multi-GEM structures has been explored.
The numerical and geometrical details of the primary seed have been found to have significant effect on the resulting probability of having either avalanche, or discharge mode operation.
Once again, it has been possible to mimic the experimentally observed trends for this critically important factor, despite drastic simplifying assumptions.
Finding an optimum point in ($\frac{\Delta{V}}{V}$) for improvement in the discharge rates in an experiment can become a challenge as the discharge formation and propagation is a complex process that depend on tuning of other several performance parameters of the detector as well. 
The present fast hybrid model may be of some help to predict the discharge rates in an experimental scenario utilizing an asymmetric configuration of operating voltages in a multistage GEM-based device with reasonable confidence.  

\acknowledgments
This work has been performed in the framework of RD51 collaboration. We wish to acknowledge the members of the RD51 collaboration for their help and suggestions. 
We would like to acknowledge necessary help and support from SINP. We would also like to thank the respective funding agencies, DAE and INO collaboration. 
Author P.Bhattacharya acknowledges the University Grant Commission and Dr. D.S.Kothari Post Doctoral Scheme for the necessary support.
Finally, we would like to thank the Referees of this paper for their detailed and encouraging review, and raising very important points that allowed us to improve the manuscript in many ways.

\end{document}